\documentclass[aps,pra,showpacs,twoside,twocolumn,longbibliography,10pt]{revtex4-2}
\usepackage[colorlinks=true, citecolor=blue, urlcolor=blue, linkcolor = blue ]{hyperref}
\usepackage{epsfig,newlfont,amssymb,amsfonts,amsmath,bm,subfigure,palatino,mathtools,amsthm,braket,times,soul,enumitem,color}
\usepackage[normalem]{ulem}
\newcommand{\stkout}[1]{\ifmmode\text{\sout{\ensuremath{#1}}}\else\sout{#1}\fi}
\usepackage[english]{babel}
\usepackage[utf8]{inputenc}
\usepackage{array}
\usepackage{xcolor}
\usepackage{graphics}

\def\Tr{\text{Tr}}

\usepackage{amsthm}
\usepackage{verbatim}
\usepackage{bbm}
\usepackage{wrapfig}
\usepackage{cancel}

\usepackage{hyphenat}

\usepackage{float}

\usepackage{orcidlink}

\newlength\figureheight 
\newlength\figurewidth 

\newcommand{\mP}[0]{{\mathcal{P}}}

\newcommand{\dir}[0]{{\text{dir}}}
\newcommand{\prbe}[0]{{\text{probe}}}

\begin{document}

%\title{Modified Exchange Fluctuation Theorems for non-Markovian baths in the collisional model}

\title{Exchange Fluctuation Theorems for Non-Markovian Baths in Quantum Collisional Model}

\author{Sayan Mondal~\orcidlink{0000-0003-2921-403X}}
\email{sayanmondal96sbs@gmail.com}

\author{Sukrut Mondkar~\orcidlink{0000-0001-9212-4366}} 
\email{sukrutmondkar@gmail.com}

\author{Ujjwal Sen~\orcidlink{0000-0002-0091-5847}}
\email{ujjwal@hri.res.in, ujjwalsen0601@gmail.com}

\affiliation{Harish-Chandra Research Institute, Chhatnag Road, Jhunsi, Prayagraj  211 019, India\\
Homi Bhabha National Institute, Training School Complex, Anushakti Nagar, Mumbai
400 094, India}

\begin{abstract}
The quantum exchange fluctuation theorem relates the probabilities of observing heat transfer along and against the temperature gradient between thermal baths at different temperatures. We investigate how this relation generalizes when the baths exhibit non-Markovian dynamics. Using a microscopic collisional model, bath memory is generated through interactions between successive bath auxiliaries before each heat-exchange collision. We derive exchange fluctuation theorems for both direct bath-bath interactions and probe-mediated heat exchange in the steady-state regime. As an illustrative example, we consider heat baths with qubit auxiliaries and show that non-Markovian memory enhances the probability of heat-transfer events against the temperature gradient, modifying the predictions made by the conventional Jarzynski-W{\'o}jcik exchange fluctuation theorem. Our results establish a microscopic connection between environmental memory and non-equilibrium heat-exchange statistics.
\end{abstract}

\maketitle

\section{Introduction}

The rapid development of quantum technologies has generated a growing need to understand thermodynamics at the quantum scale~\cite{Gemmer2004, Binderbook, Deffnerbook}. This has led to extensive studies of heat and work in non-equilibrium quantum thermodynamics~\cite{Strasbergbook, Seifertbook}, with applications ranging from quantum batteries~\cite{Alicki2013, BatteriesRMP} and heat engines~\cite{Alicki1979, Kosloff1984, Quan2007, Cangemi2024} to quantum refrigerators~\cite{Linden2010, Skrzypczyk2011, Levy2012, Mitchison2019}. Since the operation of these devices relies fundamentally on the exchange of energy, understanding the statistics of heat exchange is of central importance~\cite{Esposito2009, Campisi2011}. Moreover, heat transport plays a fundamental role in the second law of thermodynamics. At the microscopic scale, thermodynamic quantities such as heat and work are inherently stochastic due to thermal and quantum fluctuations, making their full probability distributions, rather than only their average values, essential for describing non-equilibrium processes~\cite{Esposito2009, Campisi2011, Manzano2022}.

Fluctuation theorems~\cite{Evans2002,Seifert_2012} provide exact relations governing the statistics of thermodynamic quantities far from equilibrium and constitute one of the most important generalizations of the second law of thermodynamics. Prominent examples include the Jarzynski equality~\cite{Jarzynski1997} and the Crooks fluctuation theorem~\cite{Crooks1999}, which characterize the statistics of work performed during non-equilibrium processes. For heat exchange between two systems prepared at different temperatures, the corresponding relation is provided by the Jarzynski-W{\'o}jcik exchange fluctuation theorem (JW-XFT)~\cite{Jarzynski2004}. The JW-XFT relates the probabilities of heat exchange
along (forward trajectory) and against (reverse trajectory) the temperature gradient between the two baths and recovers the Clausius inequality as its consequence.

% In quantum systems, the fluctuation theorems are traditionally derived using the two-point measurement (TPM) protocol
In quantum systems, fluctuation theorems are commonly derived using the two-point measurement (TPM) protocol~\cite{kurchan2000,tasaki2000,Talkner2007,Talkner_2007JPA}, where projective energy measurements are performed before and after the dynamics. Since the initial measurement destroys quantum coherence and can modify other quantum correlations, several alternative formulations have been developed, including
% in which projective energy measurements are performed before and after the dynamics. Although the TPM protocol provides a consistent trajectory-based description of heat exchange, the initial measurement destroys quantum coherence and may alter other quantum correlations present in the initial state. This limitation has motivated several alternative formulations based on 
end-point measurements (EPM)~\cite{Gherardini2021,HernndezGmez2023,Gianani2023, Mondkar2025,Artini2026}, quasi-probability distributions~\cite{Lostaglio2018, Kwon2019, Levy2020,Landi2024,Li2025SciAdv,Jae2026}, Bayesian networks~\cite{Micadei2020, Micadei2021}, and related approaches, allowing the influence of quantum resources on fluctuation relations to be investigated. 
Over the years, exchange fluctuation theorems have also been generalized to finite baths~\cite{Akagawa2009}, correlated initial states~\cite{Jevtic2015}, strong system-bath coupling~\cite{Nicolin2011,Sone2023,Wu2024}, squeezed reservoirs~\cite{Manzano2016,Yadalam2022,Sarmah2023,Li2025,Hernandez-Gomez2025}, and systems with multiple conserved quantities~\cite{Timpanaro2019,Upadhyaya2024,Rodrigues2025,Scandi2026}.

% Most exchange fluctuation theorems assume Markovian environments, 
% \textcolor{orange}{where successive heat-exchange events are statistically independent} \textcolor{red}{[Aren't the ``heat-exchange events'' specific to collisional model environments?]}. 
{
% \color{green}
Most exchange fluctuation theorems assume a Markovian environment, so that the statistics of heat exchange are independent of the system’s earlier interaction history.
In realistic quantum systems, however, memory effects naturally arise from structured reservoirs~\cite{Mazzola2009}, finite environments~\cite{Sampaio2017}, strong system-environment coupling~\cite{deVega2017}, or bath correlations~\cite{Lorenzo2011}. Although fluctuation theorems have been studied in certain non-Markovian settings~\cite{Wu2024}, the influence of bath memory on exchange fluctuation theorems remains largely unexplored.

Quantum collisional models~\cite{Ciccarello2022,Lacroix2025} provide a microscopic framework for open-system dynamics through repeated system-environment interactions. By introducing interactions between successive bath auxiliaries~\cite{Ciccarello2013}, they naturally describe non-Markovian dynamics  
making them particularly well suited for studying the effects of non-Markovianity on fluctuation theorems.

% Despite these advances, the role of 
% \textcolor{orange}{dynamically generated }
% bath memory 
% in exchange fluctuation theorems remains largely unexplored. 
{Existing non-Markovian fluctuation theorems like Ref.~\cite{Wu2024}}, 
% \textcolor{red}{[Are there existing non-Markovian fluctuation theorems? If yes, please cite]} 
typically consider specific open-system models and strong system-environment coupling. An exact exchange fluctuation theorem for microscopic collisional models with bath-memory effects has, to our knowledge, not been established. Addressing this question is the primary motivation of the present work.
In this work, we close this gap by deriving a generalized XFT for two physically relevant scenarios: direct bath-bath interactions and probe-mediated heat exchange. We obtain exact trajectory-dependent additional factors arising from bath memory, recover the standard JW-XFT in the Markovian limit, and show quantitatively how non-Markovianity enhances the probability of {reverse heat-transfer events for an example of baths with qubit auxiliaries}. }
% \textcolor{red}{[Is saying `` reverse heat-transfer events'' enough ? or should we be more specific? e.g., see the abstract of Ref~\cite{Scandi2026}, they have written ``inversion of all currents against
% their affinity biases''....so in our case heat transfer against temperature gradient? or from low to high temperature?...we can say this once and from next occurrence of it say ``reverse heat-transfer''??]}}

The paper is organized as follows. In Sec.~\ref{Sec2}, we briefly review the standard JW-XFT, quantum collisional models, and the measure of non-Markovianity employed in this work. In Sec.~\ref{Sec3}, we derive the generalized exchange fluctuation theorems for direct bath-bath interactions and probe-mediated heat exchange, and illustrate the results through explicit numerical examples. Finally, we conclude in Sec.~\ref{Sec4}.

\section{Preliminaries}
\label{Sec2}
\subsection{Basics of standard XFTs}
 Exchange fluctuation theorems (XFTs) relate the probability of exchanging conserved quantities, such as energy or particle number, between two systems initially prepared under different thermodynamic conditions to the probability of the corresponding exchange in the time-reversed process. At the trajectory level, fluctuations may lead to exchanges that occur against the thermodynamic affinity. However, upon averaging over many realizations, the net flow of conserved quantities follow the direction imposed by the corresponding thermodynamic gradient; for instance, heat flows on average from a hotter system to a colder one, while particles flow from higher to lower chemical potential. In this sense, XFTs provide a generalization of the second law of thermodynamics to non-equilibrium processes. One of the most celebrated examples is the JW-XFT~\cite{Jarzynski2004}, originally derived for heat exchange between two systems prepared at different temperatures and shown to hold for both classical and quantum dynamics.

 % {\color{cyan} Write a bit more, how 2nd law comes.}\textcolor{red}{[I think a reader who is not familiar with quantum stochastic thermodynamics won't understand why the previous sentence implies that XFTs constitute a generalization of the second law. Maybe fill in some sentences to make a logical connection?]}
 % \textcolor{red}{[I am not able to understand the meaning of this sentence. Isn't heat exchange between two systems a dynamics? Originally derived for classical or quantum systems? I guess for both from the title of the JW-XFT paper~\cite{Jarzynski2004} cited here?]}

Consider two thermal baths $A$ and $B$, initially prepared at inverse temperatures $\beta_A$ and $\beta_B$, respectively, and allowed to exchange heat through an energy-conserving interaction. The JW-XFT states that the ratio of the probability of observing a heat transfer $q$ from bath $B$ to bath $A$ to that of observing the reverse transfer $-q$ is given by
\begin{align}\label{JW-XFT}
    \frac{p(q)}{p(-q)} = e^{\Delta\beta\,q},
\end{align}
where
\begin{align}
    \Delta\beta={\beta_A-\beta_B.}
\end{align}

For the case $T_B>T_A$ (equivalently $\beta_B<\beta_A$), we have $\Delta\beta>0$, implying that a heat flow from the hotter bath $B$ to the colder bath $A$ is exponentially more probable than the reverse process.

Summing the fluctuation relation, {Eq.~\eqref{JW-XFT},} over all possible heat exchanges yields the integral fluctuation theorem,
\begin{align*}
    \sum_q p(q)e^{-\Delta\beta q}=\sum_q p(-q)=1,
\end{align*}
or equivalently, $\left\langle e^{-\Delta\beta q}\right\rangle = 1$.
Applying Jensen's inequality then gives
\begin{align*}
    \Delta\beta\,\langle q\rangle \ge 0,
\end{align*}
which is precisely the Clausius form of the second law, stating that the average heat flows spontaneously from the hotter system to the colder one.

Since its original formulation, the JW-XFT has been generalized in several directions. These include heat currents in open quantum systems~\cite{Andrieux_2009}, finite thermal reservoirs~\cite{Akagawa2009}, strong system-bath coupling in spin-boson and oscillator-reservoir models~\cite{Nicolin2011,Wu2024,Denzler2018}, initial correlations~\cite{Jevtic2015,Sone2023}, and non-thermal initial states characterized by an effective inverse temperature~\cite{Hernandez-Gomez2025}. Exchange fluctuation relations have also been investigated for non-classical environments, including squeezed thermal reservoirs and their applications to quantum thermal machines~\cite{Manzano2016}, bosonic transport~\cite{Yadalam2022,Sarmah2023}, and full counting statistics in squeezed environments~\cite{Li2025}. Related studies have examined the role of squeezing and coherence using the Margenau-Hill quasiprobability formalism~\cite{PhysRevE.108.054109}, classical Brownian transducers and non-equilibrium steady states~\cite{Gomez-Marin2006,PhysRevE.103.042143}, and simplified formulations for bosonic baths~\cite{Devi2021}.

More recently, XFTs have been extended to systems with multiple conserved quantities and generalized Gibbs ensembles~\cite{Wei2018, Manzano2022PRXQ}, establishing connections with thermodynamic uncertainty relations~\cite{Timpanaro2019,Hasegawa2019}, non-Abelian charge transport~\cite{Scandi2026}, Bayesian-network formulations~\cite{Rodrigues2025}, alternative notions of entropy production~\cite{Upadhyaya2024}, and corresponding TURs~\cite{Salazar:2025qfs}. 
Other developments include anomalous transport induced by initial correlations~\cite{gy6n-5x26}. 
% Comprehensive reviews of exchange fluctuation theorems can be found in Refs.~\cite{Esposito2009, Campisi2011}.
\subsection{Quantum collisional model{s}}
The quantum collisional model~\cite{Ciccarello2022} is a framework that is used to study open quantum systems and quantum thermodynamics. Collisional models 
{provide} a microscopic description of open quantum system dynamics in which the environment is represented by a sequence of auxiliary {units} that interact with the system one at a time. 

The environment consists of a sequence of auxiliary {systems} $\{E_n\}_{n=1}^{N}$, each described by the same Hilbert space $\mathcal{H}_E$. The initial state of the environment is assumed to be
$\rho_E(0)=\bigotimes_{n=1}^{N}\gamma_n$,
where $\gamma_n$ denotes the initial state of the $n$-th auxiliary.
For thermal environments, all auxiliaries are identically prepared in the Gibbs state, $\gamma_n={e^{-\beta H_E}}/{\Tr(e^{-\beta H_E})}$.
The system interacts sequentially with the auxiliaries. During the $n$-th collision, the joint evolution of the system and the corresponding auxiliary is governed by the unitary operator
$U_n=e^{-(i/\hbar)H_n\tau}$,
where 
% \textcolor{orange}
% \textcolor{cyan}
{$H_n$ is the total Hamiltonian acting on the system and auxiliary, describing the collision} 
% \textcolor{red}{[Do we need to specify what $H_n$ is in terms of system, auxiliary, and interaction terms?]} 
and $\tau$ is the interaction time. The post-collision state is given by
$\rho'_{SE_n} = U_n(\rho_S\otimes\gamma_n)U_n^\dagger$.
The reduced state of the system is then obtained by tracing over the bath auxiliary.

In the standard collisional model, each auxiliary interacts with the system only once before being discarded. Consequently, successive collisions are statistically independent, giving rise to Markovian dynamics. Memory effects are incorporated by allowing additional interactions between auxiliaries, by preparing the auxiliaries in correlated initial states, or by allowing the
system to interact multiple times with the same auxiliary. These mechanisms generate correlations within the environment, which can subsequently influence future system dynamics, leading to non-Markovian evolution.
In this work, we introduce non-Markovianity by considering 
auxiliary-auxiliary collisions. 

\subsection{Non-Markovianity}
An open quantum system is said to exhibit Markovian dynamics when its evolution is memoryless, i.e., the future evolution of the system depends only on its present state and not on its past history. In contrast, non-Markovian dynamics arise when the environment retains memory of previous system-environment interactions, leading to information backflow from the environment to the system. 
% {\color{cyan}\sout{Such memory effects may originate from strong system-environment coupling, structured environments, or environmental correlations and are known to significantly influence quantum thermodynamic processes.} }\textcolor{red}{[CITE for significantly influence quantum thermodynamic processes]}

Several measures of non-Markovianity exist in literature. In this work we focus on the Breuer-Laine-Piilo (BLP) measure~\cite{Breuer2009, Laine2010} of non-Markovianity. This is based on the rate of change of trace distance between two quantum states. The trace distance between any two quantum states $\rho$ and $\sigma$ is given by, $D(\rho,\sigma) = \frac{1}{2}|\rho-\sigma|$,
where $|A| = \Tr[\sqrt{A^\dagger A}]$. The BLP measure is given by,
\begin{align}
    \mathcal{N}^c = \max_{\rho(0),\sigma(0)}\int_{\partial_t D > 0}\partial_t D(\rho(t), \sigma(t)) dt.
\end{align}
Here, in the integral, only time intervals where this rate of trace distance is positive is included. Moreover, we have an optimization over all possible pairs of initial states. In the current work, we consider the collisional model with discrete time steps. Hence, we use a discretized version of the BLP measure~\cite{McCloskey2014, Senyasa2022,  McElvogue2026} given by
\begin{align}
    \mathcal{N} = \max_{\rho_0,\sigma_0} \sum_n \left[ D(\rho_n,\sigma_n) - D(\rho_{n-1}-\sigma_{n-1})\right].
    \label{eq:NM}
\end{align}
Here again, in the sum we only add positive values of the summand, and $\rho_n$ and $\sigma_n$ are the respective states after the $n$-th collision corresponding to the initial states $\rho_0$ and $\sigma_0$ respectively. Since the trace distance $D(\rho,\sigma)$ is dimensionless, the BLP measure $\mathcal{N}$ is also a dimensionless quantity.
% {\color{cyan}Non-Markovianity studied in~\cite{Zhang2021, McElvogue2026}.}
\section{Exchange fluctuation theorems generalized to non-Markovian baths}
\label{Sec3}
% In this section, we present the main results of our work. We investigate the modified XFTs in the presence of non-Markovian baths realized through the collisional model. 
The derivation presented here constitutes the central result of this work. Unlike previous formulations of the {XFTs}, which assume Markovian environments, we derive an exact modification arising from 
% \textcolor{cyan}{\sout{dynamically generated} 
{bath memory} within a microscopic collisional model. The resulting correction factor depends solely on the non-Markovian correlations generated between successive bath auxiliaries.

\begin{figure*}[t]
    \centering
    \includegraphics[width=0.9\linewidth]{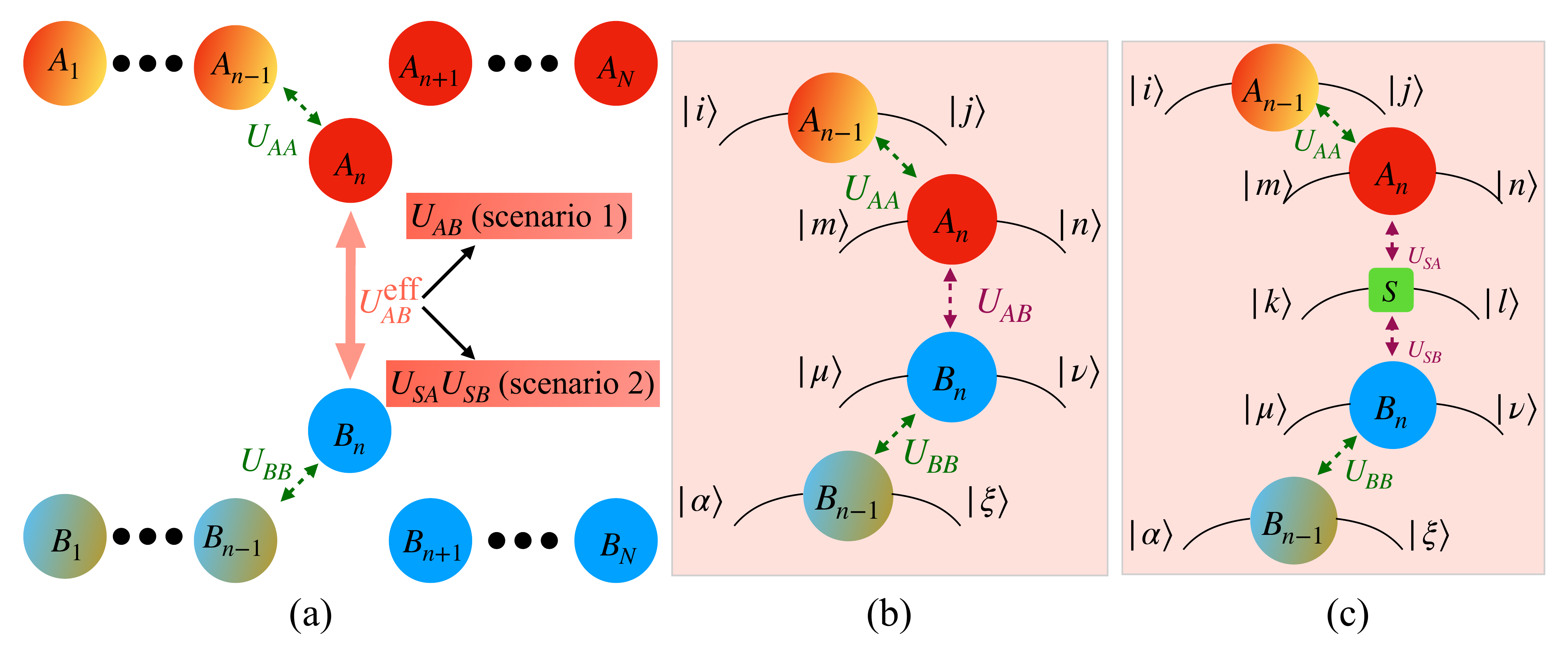}
  \caption{\textbf{Schematic diagram of the setup.} (a) Collisional model describing heat exchange between two baths $A$ and $B$. The red (blue) circles denote auxiliary units of bath $A$ ($B$), each initially prepared in a Gibbs state at the corresponding bath temperature. 
  Before the $n$th inter-bath collision, the neighboring auxiliaries within each bath undergo intra-bath interactions $U_{AA}$ and $U_{BB}$, generating memory effects.
  The intra-bath interactions involve the $n$th and $(n-1)$th auxiliary unit of each bath. The initial state of the $n$th unit is the thermal Gibbs state, while that of the $(n-1)$th unit can be some athermal state.
  The subsequent effective interaction between the two baths is denoted by $U_{AB}^{\text{eff}}$. We consider two realizations: (i) direct bath-bath interaction, $U_{AB}^{\text{eff}}=U_{AB}$, and (ii) probe-mediated interaction, $U_{AB}^{\text{eff}}=U_{SA}U_{SB}$. Furthermore, we present the TPM protocol for (b)the direct interaction case, and (c)the probe-mediated interaction. In (c), an intermediate probe system $S$ sequentially interacts with baths $A$ and $B$ through $U_{SA}$ and $U_{SB}$.}
    \label{fig1}
\end{figure*}

\subsection{General framework}

The standard formulations of XFTs typically assume Markovian thermal baths. 
% \textcolor{cyan}{\sout{where successive interactions are statistically independent}}. \textcolor{red}{[Phrase ``successive interactions'' is specific to the context of collisional model baths, or is it applicable for any bath in general? If it is the former, then saying ``standard formulations of XFTs assume Markovian thermal baths, where successive interactions are statistically independent'' is a bit inaccurate??]} 
However, realistic environments often possess memory effects and induce non-Markovian dynamics, leading to correlations between different stages of the heat exchange process.

Collisional models provide a natural framework for studying such questions. Besides offering a microscopic description of open-system dynamics, they also allow a transparent implementation of the two-point measurement (TPM) protocol, which is commonly employed in deriving fluctuation theorems~\cite{Santos2020, Schmidt2023, Scandi2026}. 
% In Ref.~\cite{Scandi2026} collisional model was used to find the XFT for systems with non-Abelian conserved charges, but with the baths being assumed to be Markovian in nature. In Ref.~\cite{Santos2020}, the authors investigated heat exchange in a collisional model, where a quantum system undergoes sequential collisions with an arbitrary number of environmental auxiliaries, focusing on the joint probability distribution for the exchanged heats with each auxiliary. {\color{cyan} The authors generalize the JW-XFT with the system interacts sequentially with multiple reservoirs at different temperatures. In the current work, we generalize the JW-XFT, considering the effects of non-Markovianity arising from the auxiliary-auxiliary collision.}
In Ref.~\cite{Scandi2026}, a collisional model was employed to derive a XFT for systems with non-Abelian conserved charges, under the assumption that the baths are Markovian. In Ref.~\cite{Santos2020}, the authors studied heat exchange in a collisional setting where a quantum system interacts sequentially with an arbitrary number of environmental auxiliaries, and analyzed the joint probability distribution of the heat exchanged with each auxiliary. {In particular, they generalized the JW-XFT to the case in which the system interacts sequentially with multiple Markovian reservoirs at different temperatures. In the present work, we extend the JW-XFT in a complementary direction by incorporating non-Markovian effects generated by collisions between the environmental auxiliaries themselves.}
% \textcolor{red}{[Should we also clarify why our work is significantly different from Ref.~\cite{Santos2020}? Maybe identify limitations of Ref.~\cite{Santos2020} or mention things that Ref.~\cite{Santos2020} didn't address and then say that our work makes progress by overcoming that limitation of Ref.~\cite{Santos2020} or addressing novel things that Ref.~\cite{Santos2020} didn't address?]}

% In this work, we go beyond the Markovian paradigm and investigate the modifications to the XFTs for heat exchange arising from non-Markovian bath dynamics. 
In the present work, we model the non-Markovianity through interactions between auxiliary units belonging to the same bath, thereby generating memory effects and correlations across successive collisions. 
The derivation presented below is independent of the specific form of the interaction Hamiltonians and relies only on the collisional structure and the assumption of global energy conservation.
We show that these memory effects lead to trajectory-dependent corrections to the standard XFT. The modified XFT obtained reduces to the conventional form in the Markovian limit. 

{
% \color{cyan}
Let us consider two baths, $A$ and $B$, initially prepared in thermal equilibrium at temperatures $T_A$ and $T_B$, respectively. Thus, initially, every constituent unit of each bath is prepared in the Gibbs state corresponding to temperature $T_{A/B}$. The dynamics involve two types of interactions: the inter-bath interaction, represented by $U_{AB}^\text{eff}$, and the intra-bath interactions within baths $A$ and $B$, represented by $U_{AA}$ and $U_{BB}$, respectively. 
The auxiliary units of each bath first interact among each other, i.e. the $(n-1)$th and $n$th unit of baths $A$ and $B$ interact with each other through $U_{AA}$ and $U_{BB}$, respectively.
After this, the $n$th units of baths $A$ and $B$ interact with each other under $U_{AB}^\text{eff}$.
% After each intra-bath collision implemented by $U_{AA}$ and $U_{BB}$, the involved particles 
% \textcolor{orange}{subsequently interact with another particle belonging to the other bath} \textcolor{red}{[can be rephrased for clarity?]} through $U_{AB}^\text{eff}$. 
These repeated intra-bath collisions propagate information through the bath and induce memory effects in the reduced dynamics. 
The degree of non-Markovianity is therefore controlled by the intra-bath interactions. In particular, in the limit $U_{AA}=I$ (or $U_{BB}=I$), bath $A$ (or $B$) reduces to a Markovian bath, since no memory is transferred between successive collisions. We present this setup schematically in Fig.~\ref{fig1}(a).

{The reduced dynamics of the colliding auxiliary units can be described by a completely positive trace-preserving (CPTP) map, $\Lambda_{n-1\to n}$, which transforms the joint state of the $(n-1)$-th auxiliary pair into that of the $n$-th auxiliary pair. Explicitly,
\begin{align}
    \rho_{A_nB_n}=\Lambda_{n-1\to n}\!\left(\rho_{A_{n-1}B_{n-1}}\right),
\end{align}
where
$\Lambda_{n-1\to n}(\rho_{A_{n-1}B_{n-1}}) = \Tr_{A_{n-1}B_{n-1}}[\mathcal{U}(\rho_{A_{n-1}B_{n-1}}\otimes \gamma_{A_nB_n})\mathcal{U}^\dagger] $.
Here, $\gamma_{A_nB_n}=\gamma_{A_n}^{\beta_A}\otimes\gamma_{B_n}^{\beta_B}$ denotes the fresh auxiliary pair prepared in thermal equilibrium, and $\mathcal{U}=U_{AB}^{\text{eff}}(U_{AA}\otimes U_{BB})$ is the unitary describing one complete collision cycle, including both the inter-bath and the intra-bath collisions.

Since every collision cycle involves identical interaction unitaries and freshly injected thermal auxiliary units, the map $\Lambda_{n-1\to n}$ is independent of the collision step. Furthermore, as a CPTP map acting on a finite-dimensional Hilbert space, it admits at least one fixed point~\cite{Scarani2002,Ziman2002,Wolf2012,Saha2024}. Consequently, after sufficiently many collision steps, the state of the colliding auxiliary pair converges to a steady state satisfying
\begin{align}
\Lambda_{n-1\to n}\left(\sigma^{\text{ss}}_{AB}\right)=
\sigma^{\text{ss}}_{AB},
\end{align}
where $\sigma^{\text{ss}}_{AB}$ denotes the steady-state density matrix.

It is important to note that the existence of the CPTP map for the auxiliary pairs does not imply that the individual baths evolve under Markovian dynamics. The non-Markovian character of the bath dynamics arises from the intra-bath interactions, through which previously interacted auxiliary units carry memory of earlier collisions. Nevertheless, at each collision step, the outgoing auxiliary pair interacts only with freshly prepared thermal auxiliary units.   
Subsequently, the two pairs are in a product state and evolve under a combined unitary in the current collisional cycle.
{Thus, the evolution of the colliding auxiliary pairs admits a Markovian description in the joint state space}, even though the reduced dynamics of each bath remains non-Markovian.}
}

Although the following derivation of the XFT does not require the system to attain a steady state, in this work we focus on the steady-state regime of the dynamics. 
The same procedure can be used to derive the transient XFT, for which the state of the auxiliary units before each collision is required.
To derive the modification to the XFT arising from non-Markovian effects, we employ the two-point measurement (TPM) protocol at each collision step. 
% Throughout our analysis, we assume that the dynamics have reached the steady-state regime. 
In the steady-state regime, the process becomes effectively cyclic: the final state of the auxiliary units $A_n$ and $B_n$ after the $n$-th collision step serves as the initial state for the subsequent $(n+1)$-th step, where they interact with $A_{n+1}$ and $B_{n+1}$, respectively.

Thus, let us consider the $n$-th collision step. At the beginning of the process, projective energy measurements are performed on the particles $A_{n-1}$, $A_n$, $B_{n-1}$, and $B_n$. In the case of probe-mediated bath-bath interaction, the energy of the probe is also measured. The corresponding joint eigenstate is denoted by $\ket{\bar{\mathbf{i}}}$. The particles then evolve under the global energy-conserving unitary
$\mathcal{U}$. Following the evolution, a second projective energy measurement is performed, yielding the final state $\ket{\bar{\mathbf{f}}}$. This defines the forward trajectory
$\Gamma:\ket{\bar{\mathbf{i}}}\rightarrow\ket{\bar{\mathbf{f}}}$.
Similarly, we define the reverse trajectory
$\tilde{\Gamma}:\ket{\bar{\mathbf{f}}}\rightarrow\ket{\bar{\mathbf{i}}}$,
generated under the time-reversed evolution operator
$\tilde{\mathcal{U}}=\Theta\mathcal{U}^{\dagger}\Theta^{-1}$.

Using these forward and backward trajectories, the generalized XFT
% \textcolor{orange}{exchange fluctuation theorem} \textcolor{red}{[write ``exchange fluctuation theorem'' only at the first instance and write XFT at all the subsequent instances?]} 
takes the form
\begin{align}
\frac{{\mP}(Q)}{{\mP}_B(-Q)} = e^{\Delta\beta Q}\,\mathcal{C},
\label{XFT-general}
\end{align}
where $\Delta\beta=\beta_A-\beta_B$, and $\mathcal{C}$ is an additional factor arising from the non-Markovian nature of the baths. In the Markovian limit, corresponding to heat exchange between two memoryless baths, $\mathcal{C}=1$, and Eq.~\eqref{XFT-general} reduces to the standard 
% \textcolor{orange}{exchange fluctuation theorem} \textcolor{red}
{form of JW-XFT}. Here, $\mP(Q)$ denotes the probability that an amount of heat $Q$ is transferred from bath $B$ to bath $A$ during the forward process, while $\mP_B(-Q)$ denotes the probability of observing the corresponding reverse heat transfer, $-Q$, during the backward process.

Summing both sides of Eq.~\eqref{XFT-general} over all possible heat exchanges yields the integral 
XFT,
% \textcolor{orange}{exchange fluctuation theorem},
\begin{align}
\sum_Q \mP(Q)e^{-\Delta\beta Q-\ln\mathcal{C}} &= \sum_Q \mP_B(-Q)=1,\nonumber\\
\Big\langle e^{-\Delta\beta Q-\ln\mathcal{C}}\Big\rangle &= 1,\nonumber
\end{align}
which, upon applying Jensen's inequality, gives
\begin{align}
\Delta\beta\langle Q\rangle+\langle\ln\mathcal{C}\rangle\ge0.
\label{eq:Int-XFT}
\end{align}
This relation represents the modified 
% \textcolor{orange}{stochastic} \textcolor{red}{[This gives a bound on average quantities, so can we call it stochastic?]} 
second law, where the non-Markovian contribution is captured by the additional term $\langle\ln\mathcal{C}\rangle$.

{The factor $\mathcal{C}$ quantifies the influence of bath memory on the heat-exchange statistics. In particular, $\mathcal{C}<1$, enhances the relative probability of 
% \textcolor{orange}
{heat-exchange} events against the temperature gradient, corresponding to heat flowing from the colder bath to the hotter bath, whereas $\mathcal{C}>1$ favors the forward heat flow from the hotter bath to the colder bath.}

In the following subsections, we explicitly derive the non-Markovianity-induced factor ${\mathcal{C}}$ for the two different interaction scenarios: direct bath-bath interaction and probe-mediated bath-bath interaction. For each scenario, we analyze both cases where only one bath is non-Markovian and where both baths are non-Markovian.

\subsection{Direct bath-bath interaction}
Let us now consider the case where the two bath units interact directly with each other. We explicitly derive the modified XFT for such a case. The situation is schematically presented in Fig.~\ref{fig1}(b). As discussed in the previous subsection, we perform an energy-projective measurement on the auxiliary particles $A_{n-1}$ and $A_n$ of bath $A$ and $B_{n-1}$ and $B_n$ of bath $B$. After the measurement, the state of the four particles comes out to be $\ket{\mathbf{i}} = |i_Am_A\alpha_B\mu_B\rangle$, where $i_A$ and $\mu_B$ correspond to the $n$th units, whereas $m_A$ and $\alpha_B$ correspond to the $(n-1)$th units of their respective baths.
% \textcolor{red}
% {[Maybe we should clarify this notation more specifying that first two entries are for $(n-1)$-th qubit pair and next two entries are for $n$-th qubit pair]} 
The probability of getting this state is given by
\begin{align}
    p({i,m,\alpha,\mu}) = p_m^{\beta_A} p_\mu^{\beta_B} q_{i\alpha}^{AB},
\end{align}
where $p_{m}^{\beta_{A}} = \bra{m}\gamma_{\beta_{A}} |m\rangle$ and $p_{\mu}^{\beta_{B}} = \bra{\mu}\gamma_{\beta_{B}} |\mu\rangle$.
% , is the probability of getting the state $\ket{\eta}$ after the initial projective measurement.  
% \textcolor{red}{[Shall we use different notations than $o$ and $\eta$ ? Especially for $\eta$, because Latin are reserved for $A$ and Greek are reserved for $B$, so using $\eta$ as common is a bit disruptive?]}
Similarly, $q_{i\alpha}^{AB} = \bra{i\alpha}\sigma^\text{ss}_{AB} |i\alpha\rangle$. Here, $\gamma_{\beta_{A/B}}$ is the thermal Gibbs state at inverse temperature $\beta_{A/B}$ and $\sigma^\text{ss}_{AB}$ is the steady state that the auxiliary units reach after repeated collisions. { The joint-initial state of the $(n-1)$th auxiliary units before the $n$th collision is $\sigma_{AB}^{\text{ss}}$, whereas the $n$th auxiliary units are initially in the thermal Gibbs state, $\gamma_{\beta_{A/B}}$. }
% \textcolor{red}{[Shouldn't we clarify in terms of index $n$ and $n-1$ which two units are in Gibbs state and which two combined in steady state?]}
After the measurement, the $(n{-}1)$-th particle interacts with the $n$-th particle of the same bath.  These interactions are modeled with the unitaries $U_{AA}$ and $U_{BB}$ for bath $A$ and $B$, respectively. Then the $n$-th particle of both baths interacts with each other. This is modeled with the unitary $U_{AB}$. Hence, the total unitary operation is $\mathcal{U} = U_{AB}(U_{AA}\otimes U_{BB})$. We reiterate here that all three of $U_{AA}$, $U_{BB}$, and $U_{AB}$ conserve the total energy; consequently, $\mathcal{U}$ also conserves the total energy. Thus, we have $[\mathcal{U}, H_{A_n}+H_{A_{n-1}}+H_{B_n}+H_{B_{n-1}}] = 0$.
After this, the final energy measurement is performed. This collapses the state to $\ket{\mathbf{f}} = |j_An_A\xi_B\nu_{B}\rangle$. Thus, the probability of the trajectory $\Gamma : |\mathbf{i}\rangle \rightarrow |\mathbf{f}\rangle$ is
\begin{align}
    P^{\dir}(\Gamma) = p_m^{\beta_A} p_\mu^{\beta_B} q_{i\alpha}^{AB} |\langle\mathbf{f}|\mathcal U|\mathbf{i}\rangle|^2.
    \label{eq:xft1:fortaj}
\end{align}
Now, we consider the backward process with the reverse trajectory $\tilde{\Gamma}: \ket{\mathbf{f}} \to \ket{\mathbf{i}}$. The details {of} the reverse trajectory are discussed in Appendix~\ref{AppA}.
We get the probability of the reverse trajectory to be,
\begin{align}
    P^{\dir}_B(\tilde{\Gamma}) = p_n^{\beta_A} p_\nu^{\beta_B} q_{j\xi}^{AB} |\langle\mathbf{i}|\mathcal U^\dagger|\mathbf{f}\rangle|^2.
    \label{eq:xft1:backtaj}
\end{align}
Thus, the ratio of the probability of the forward trajectory to that of the reverse trajectory is given by,
\begin{align}
    \frac{P^{\dir}(\Gamma)}{P^{\dir}_B(\tilde \Gamma)} = e^{\Delta \beta Q_\Gamma} e^{-\beta_A \Delta f_{ji}^A} e^{-\beta_B \Delta f_{\xi \alpha}^B}e^{-\Delta{I_{j\xi,i\alpha}^{AB}}}.
    \label{eq:ratio-traj-direct}
\end{align}
Here, we define $Q_{\Gamma} \coloneq E_{n} - E_{m} + E_{j} - E_{i}$ and $\Delta \beta = \beta_A - \beta_B$. The quantity $\Delta f_{kl}^{A/B} = f_k^{A/B} - f_l^{A/B}$, where $f_k^{A/B} \coloneq E_k - \beta_{A/B}^{-1}(-\ln q_k^{A/B})$ 
% \textcolor{red}{[Should define $q_k^o$ (with one subscript); As a subscript of $f^o$, we should use something that is neither Greek nor Latin so that A is unambiguously associated with Latin and B with Greek]} 
is the stochastic free energy. We note that the change in stochastic free-energy $\Delta f_{kl}^{A/B} = 0$, for all values of $k$ and $l$, when the corresponding states are thermal states, i.e. $q_{k/l}^{A/B} \propto \exp[-\beta_{A/B}{E_{k/l}}]$. Thus, a non-zero contribution arises from the athermal nature of the corresponding state.
% , with $o\in \{A,B\}$. 
Moreover, $\Delta I_{j\xi,i\alpha}^{AB} = I^{AB}_{j\xi} - I_{i\alpha}^{AB}$, with  $I_{xy}^{AB} = \ln\left({q^{AB}_{xy}}/{q^A_xq^B_y}\right)$ is the stochastic mutual information. 
 
Moreover, we have
\begin{align}
\mP^{\dir}(Q) &= \sum_{\Gamma} P^{\dir}(\Gamma) \delta(Q_\Gamma - Q),\nonumber \\
\mP^{\dir}_B(-Q) &= \sum_{\tilde \Gamma} P^{\dir}_B(\tilde \Gamma)\delta(Q_{\tilde \Gamma} + Q).
\label{eq:def-Q-prob}
\end{align}
From Eqs.~\eqref{eq:ratio-traj-direct} and \eqref{eq:def-Q-prob}, we get 
\begin{align}
    \frac{\mP^{\dir}(Q)}{\mP^{\dir}_B(-Q)} = e^{\Delta \beta Q} \mathcal{C}^{\dir}_{\text{J}_A-\text{J}_B}.
    \label{eq:direct-XFT1}
\end{align}
Here, the superscript $\text{J}_\chi$ signifies whether the bath is Markovian or non-Markovian, viz. $\text{J}_\chi \in \{\text{NM,M}\}$. Consequently, the additional factors corresponding to various bath-combinations, are 
\begin{align}
    \mathcal{C}^{\dir}_{\text{NM-NM}} &= \left[\langle e^{\beta_A \Delta f^q_A} e^{\beta_B \Delta f^q_B}e^{\Delta I_{AB}^q} \rangle_Q\right]^{-1}, \nonumber \\
    \mathcal{C}^{\dir}_\text{M-NM} &=  \left[\langle e^{\beta_B \Delta f^q_B}\rangle_Q\right]^{-1},  \nonumber \\
    \mathcal{C}^{\dir}_\text{M-M} &= 1. 
\end{align}
Here we suppress the trajectory-indices as $\Delta f_A^q = \Delta f_{ji}^A$,  $\Delta f_B^q = \Delta f^B_{\xi\alpha}$, and $\Delta I_{AB}^q = \Delta I_{j\xi,i\alpha}^{AB}$ . 
% \textcolor{red}{[Also write for $\Delta I_{AB}^q = \Delta I_{j\xi,i\alpha}^{AB}$]}
Moreover, the conditional average $\langle X\rangle_Q$ is given by
\begin{align}
\langle X\rangle_Q = \frac{\sum_\Gamma P(\Gamma)X\delta(Q_\Gamma-Q)}{\sum_\Gamma P(\Gamma)\delta(Q_\Gamma-Q)}.
\label{cond-avg}
\end{align}
When both baths are non-Markovian, the correction factor depends on the conditional average of the exponentials of the stochastic free-energy changes of the auxiliary units together with the stochastic mutual information between them. 
% \textcolor{orange}
{The free-energy contributions arise because two of the auxiliary units are no longer in local thermal equilibrium before the current collision, while the mutual-information term accounts for the correlations established during previous collisions.} 
% \textcolor{red}{[elaborate this, especially the free energy contribution origin?]} 
These two contributions together quantify the deviation from the standard Markovian XFT.
% \textcolor{orange}{exchange fluctuation theorem}. 
In the special case where only one bath is non-Markovian, only the corresponding stochastic free-energy correction survives, whereas in the fully Markovian limit all correction terms vanish, yielding $\mathcal{C}^{\dir}_{\text{M-M}}=1$. 
% \textcolor{orange}
{We note that such quantum generalizations to fundamental thermodynamic relations involving changes in free-energy and mutual information have been reported previously in various contexts~\cite{Esposito2010, PhysRevE.99.012120, jiang2018improved, Bera2017, mondal2023modified, Aimet2025}. Here, these quantities emerge naturally as trajectory-dependent contributions to the XFT due to bath memory effects in the collisional model.} 
% \textcolor{red}{[The last two sentences sound very ChatGPT, especially the first of the two]}

% \subsubsection*{Example: direct heat exchange}
\emph{\textbf{Example.}} 
We now illustrate our results using a model in which the auxiliary units of both baths are two-level systems with local Hamiltonians $H_{A_n} = H_{B_n} = \omega|1\rangle \langle 1 |$, for every auxiliary unit $n$. The intra-bath collisions are governed by $U_{AA}$ and $U_{BB}$ for bath $A$ and $B$ respectively. We have $U_{AA} = e^{i\kappa_AH_{AA}}$ and $U_{BB} = e^{i\kappa_BH_{BB}}$, with $H_{AA} = H^{+-}_{A_nA_{n-1}}$, $H_{BB} = H^{+-}_{B_nB_{n-1}}$, and  $H_{op}^{+-} = \sigma^+_o\sigma^-_p + \sigma^-_o\sigma^+_p$, where $\sigma_o^+ = (\sigma_o^-)^\dagger = |1\rangle\langle 0|$. 
% \textcolor{red}{[Maybe write here $\sigma^+_o$, $\sigma^-_p$ etc are Pauli matrices]} 
Similarly, $U_{AB} = e^{igH_{AB}}$, with $H_{AB} = H^{+-}_{A_nB_n}$. 
We note that the quantities $\kappa_A$, $\kappa_B$, and $g$ are dimensionless interaction parameters.  Physically, these parameters are obtained by combining the corresponding coupling strengths and collision durations with the reduced Planck's constant, $\hbar$ into a single dimensionless quantity.

We first characterize the degree of non-Markovianity induced by the intra-bath interactions using the BLP measure. 
{The BLP measure $\mathcal{N}$ of the bath B, with qubit auxiliaries, is evaluated by evolving a probe qubit $Z$ coupled to the non-Markovian bath $B$ under the collisional dynamics. }
{The qubit $Z$ undergoes interaction with the qubit auxiliary units of the bath $B$, under the unitary $U = e^{igH^{+-}_{ZB}}$. The non-Markovianity is induced due to the intra-bath collisions between the qubit auxiliaries of the bath, given by $U_{BB} = e^{i\kappa H_{BB}^{+-}}$.}
Here, $\kappa$ is a dimensionless interaction parameter characterizing the strength of the intra-bath collision. More precisely, if the underlying interaction Hamiltonian is $\tilde H_{BB}^{+-} = -J_{BB}H_{BB}^{+-}$ and the collision lasts for a time $\tau_{BB}$, then $\kappa={J_{BB}\tau_{BB}}/{\hbar}$,
where $J_{BB}$ sets the interaction-energy scale. Thus, $\kappa$ represents the dimensionless interaction angle accumulated during an intra-bath collision.

\begin{figure}[!h]
    \centering
    \includegraphics[width = 0.45\textwidth]{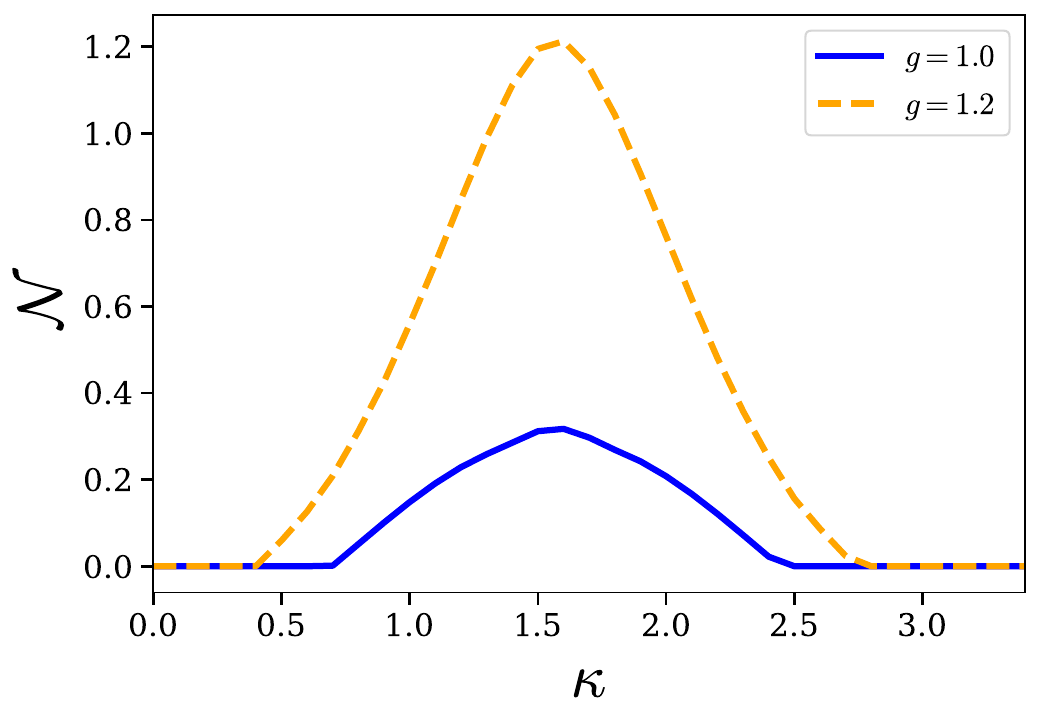}
   \caption{\textbf{BLP measure of non-Markovianity of the collisional bath.} The BLP measure of non-Markovianity, $\mathcal{N}$, is shown as a function of the intra-bath coupling strength $\kappa$. The environment consists of auxiliary qubits interacting sequentially with a system qubit. Both the system and bath auxiliary qubits have energy splitting $\omega=1.0$. Results are shown for two values of the system-bath coupling strength, $g=1.0$ (blue solid) and $g=1.2$ (orange dashed). The bath auxiliaries are initially prepared in Gibbs states with inverse temperature $\beta=1.0$. {Quantities plotted on both axes are dimensionless.}}
    \label{fig2-NM}
\end{figure}

Fig.~\ref{fig2-NM} shows the BLP measure $\mathcal{N}$ as a function of the intra-bath interaction strength $\kappa$. 
% \textcolor{red}{[$\kappa$ is not introduced before only $\kappa_A$ and $\kappa_B$ are introduced. I mean we should maybe write the interaction unitary in terms of $\kappa$? or at least say in words?]}
The non-Markovianity initially increases with $\kappa$, reaches a maximum, and subsequently decreases. Owing to the unitary nature of the intra-bath collisions, $\mathcal{N}$ is periodic in $\kappa$; the figure displays one complete period. Results are shown for two values of the inter-bath coupling: $g=1.0$ (blue solid) and $g=1.2$ (orange dashed). The analysis is done with inverse temperatures $\beta_B=1.0$. 
% \textcolor{orange}{The BLP measure is evaluated by evolving a probe qubit coupled to the non-Markovian bath under the collisional dynamics.} \textcolor{red}{Maybe move this orange sentence to the beginning of BLP discussion. I mean it is better to clarify for specifically what system the BLP is computed and after that discuss the BLP results}

\begin{figure}[!h]
    \centering
    \includegraphics[width = 0.45\textwidth]{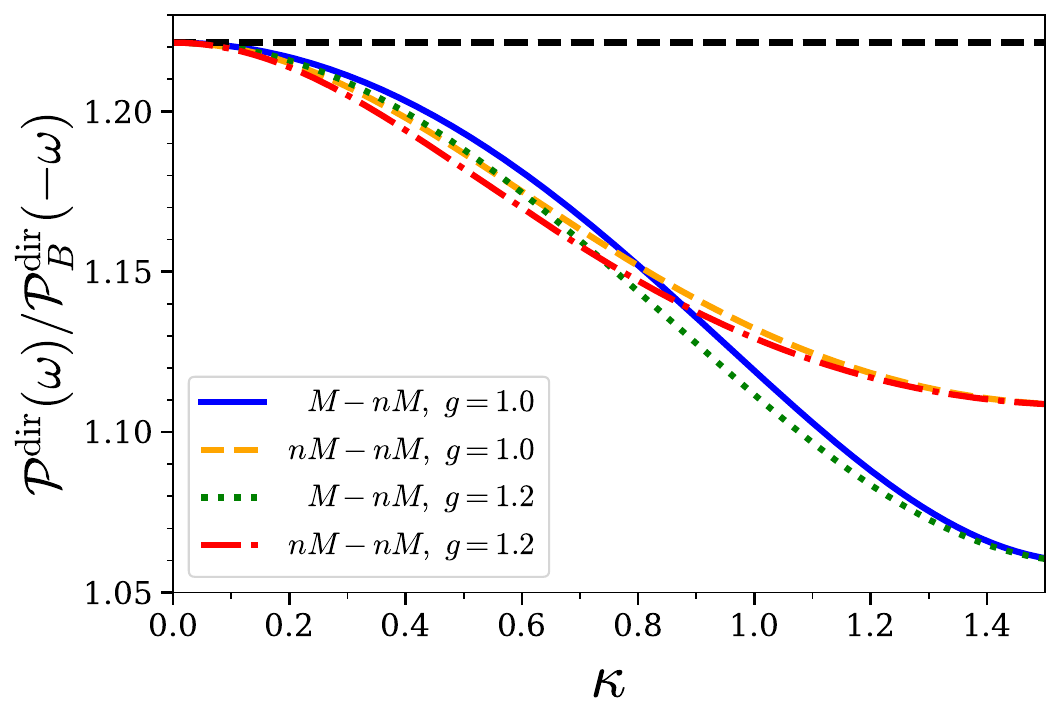}
    \caption{\textbf{XFT for direct bath-bath interaction.} The left-hand side of Eq.~\eqref{eq:direct-XFT1} is plotted as a function of the intra-bath coupling strength $\kappa$. The bath auxiliaries are qubits with energy splitting $\omega=1.0$. Results are shown for bath-bath coupling strengths $g=1.0$ and $g=1.2$, comparing a Markovian and a non-Markovian bath $A$, while bath $B$ remains non-Markovian. The inverse temperatures are $\beta_A=1.0$ and $\beta_B=0.8$. The black dashed line denotes the standard Markovian XFT prediction, $e^{\Delta\beta\omega}$, corresponding to the right-hand side of Eq.~\eqref{eq:direct-XFT1} with $\mathcal{C}^{\dir}_{\text{M-M}} = 1$. {Quantities plotted on both axes are dimensionless.}}
    \label{fig3-XFT1}
\end{figure}

Furthermore, Fig.~\ref{fig3-XFT1} shows the left-hand side of the modified XFT in Eq.~\eqref{eq:direct-XFT1} as a function of the intra-bath interaction strength $\kappa$, which controls the degree of non-Markovianity. 
We first numerically validate Eq.~\eqref{eq:direct-XFT1}, by matching the left and right-hand side of the relation.
As the non-Markovianity increases, the ratio $\mP^{\dir}(Q)/\mP_B^{\dir}(-Q)$ decreases, indicating an enhanced relative probability of 
% \textcolor{orange}
{reverse heat-transfer events}, i.e., heat flowing from the colder bath to the hotter bath. Interestingly, this suppression is more pronounced when only one bath is non-Markovian than when both baths are non-Markovian. 
% \textcolor{red}{[Do we need to speculate about the reason for this counterintuitive observation?]}

\subsection{Probe-mediated bath-bath interaction}
% \textcolor{orange}{Now we consider two baths $A$ and $B$ and a small probe system $S$. This is schematically presented in Fig.~\ref{fig1}(c). 
We now consider heat exchange between two baths, $A$ and $B$, mediated by a probe system $S$, as illustrated schematically in Fig.~\ref{fig1}(c). 
% \textcolor{red}{[Orange sentences involve repetition]} 
To derive the modified XFT for this setup, we employ the TPM protocol. Following the initial projective energy measurement, the two auxiliary units from each bath together with the probe are projected onto the joint energy eigenstate $\ket{\mathbf{I}}\coloneqq|i_Am_Ak_S\alpha_B\mu_B\rangle$ where $i_A$ and $\mu_B$ corresponds to the $n$th auxiliary units of their respective baths, $k_S$ corresponds to the probe, and $m_A$ and $\alpha_B$ corresponds to the $(n-1)$th auxiliary units of their respective baths. 
% \textcolor{red}{[Clarify order in the notation]}
The intra-bath collisions in bath $A$ and $B$ are implemented using the unitary operations $U_{AA}$ and $U_{BB}$, respectively. 
% \textcolor{red}{[Fix notation inconsistency. Fig~\ref{fig1} uses notation $U$ whereas in the text notation $V$ is used for same quantity]} 
This is followed by probe-mediated inter-bath collisions, implemented by $U_{SA}$ and $U_{SB}$.
We consider these collisions to be energy conserving. The probe first interacts with bath B, followed by bath A, so that $\mathbf{U} = (U_{SA} U_{SB})(U_{AA}\otimes U_{BB})$. After the unitary evolution, the final projective measurement is implemented with the post-measurement state being $|\mathbf{F}\rangle \coloneqq |j_An_Al_S\xi_B\nu_B\rangle$. Thus, the forward trajectory 
$\Gamma : |\mathbf{I}\rangle \rightarrow |\mathbf{F}\rangle$ is given by the probability, 
\begin{align}
    P^{\prbe}(\Gamma) = p_m^{\beta_A} p_\mu^{\beta_B} q_{ki\alpha}^{SAB}  |\langle\mathbf{F}|\mathbf {U}|\mathbf{I}\rangle|^2.
    \label{probe-for-traj}
\end{align}

Similarly, the reverse trajectory $\tilde{\Gamma} : |\mathbf{F}\rangle \to |\mathbf{I}\rangle$, is given by the probability,
\begin{align}
    P^{\prbe}_B(\tilde{\Gamma}) = p_n^{\beta_A} p_\nu^{\beta_B} q_{lj\xi}^{SAB}|\langle\mathbf{I}|\mathbf U^\dagger|\mathbf{F}\rangle|^2,
    \label{probe-back-traj}
\end{align}
where the backward protocol is constructed following the procedure outlined in Appendix~\ref{AppA}.

From Eqs.~\eqref{probe-for-traj} and \eqref{probe-back-traj}, we get 
\begin{align}
    \frac{P^{\prbe}(\Gamma)}{P^{\prbe}_B(\tilde \Gamma)} = e^{\Delta \beta Q_\Gamma} e^{-\beta_A \Delta f_{ji}^{A}} e^{-\beta_B \Delta f_{\xi \alpha}^B}e^{-\beta_B \Delta f_{lk }^S}e^{-\Delta I_{lj\xi,ki\alpha}^{SAB}}.
\end{align}
{Here, $Q_\Gamma \coloneqq E_n - E_m + E_j-E_i$, $\Delta f^A_{ji} \coloneqq f_j^A - f_i^A$, $\Delta f^B_{\xi\alpha} \coloneqq f_\xi^B - f_\alpha^B$, and $\Delta f^S_{lk} \coloneqq f^S_l - f^S_k$, with $f^S_k = E_k - \beta_B^{-1}(-\ln q_k^S)$ and $f^{A/B}_k = E_k - \beta_{A/B}^{-1}(-\ln q_k^{A/B})$. Moreover, $\Delta I_{SAB}^q =I_{lj\xi}^{SAB} - I_{ki\alpha}^{SAB}$ with $I_{x,y,z}^{SAB}=\ln(q^{SAB}_{xyz}/q_x^Sq_y^Aq_z^B)$ is the stochastic mutual-information contributions.}
In contrast to the direct bath-bath interaction, the probe-mediated protocol introduces an additional contribution arising from the non-equilibrium state of the probe. Consequently, the modified exchange fluctuation theorem takes the form
\begin{align}
    \frac{\mP^{\prbe}(Q)}{\mP^{\prbe}_B(-Q)} = e^{\Delta \beta Q} \mathcal{C}_{\text{J}_A-\text{J}_B}^{\prbe}
    \label{eq:XFT2}
\end{align}
where the correction factors are given by
\begin{align}
    \mathcal{C}_\text{NM-NM}^{\prbe} &= \left[\langle e^{\beta_B \Delta f^q_S} e^{\beta_A \Delta f^q_A} e^{\beta_B \Delta f^q_B} e^{\Delta I_{SAB}^q}\rangle_Q\right]^{-1} \nonumber \\
    \mathcal{C}^{\prbe}_\text{M-NM} &=  \left[\langle e^{\beta_B \Delta f^q_S}e^{\beta_B \Delta f^q_B}e^{\Delta I_{SB}^q}\rangle_Q\right]^{-1} \nonumber \\
    \mathcal{C}^{\prbe}_\text{M-M} &= \left[\langle e^{\beta_B \Delta f^q_S}\rangle_Q\right]^{-1}. 
\end{align}
Here, we have suppressed the trajectory-indices, such that $\Delta f_A^q$, $\Delta f_B^q$, and $\Delta f_S^q$ denote the stochastic free-energy changes of bath $A$, bath $B$, and the probe $S$, respectively. Moreover, $\Delta I_{SAB}^q =I_{lj\xi}^{SAB} - I_{ki\alpha}^{SAB}$ with $I_{x,y,z}^{SAB}=\ln(q^{SAB}_{xyz}/q_x^Sq_y^Aq_z^B)$ and $\Delta I_{SB}^q = I^{SB}_{l\xi} - I^{SB}_{k\alpha}$ are the corresponding stochastic mutual-information contributions.  The conditional average $\langle X\rangle_Q$ is defined in Eq.~\eqref{cond-avg}. Unlike the direct bath-bath interaction, the fully Markovian limit retains a nontrivial modification due to the stochastic free-energy contribution of the probe. This originates from the non-equilibrium steady state of the probe established through its repeated interactions with the baths, whenever the probe reaches a non-equilibrium steady state.

% \subsubsection*{Example: probe-mediated heat exchange}
\emph{\textbf{Example.}} 
Now, we consider an example of the 
% \textcolor{orange}
{generalized XFT} in the probe-mediated collisions case. The probe as well as the auxiliary units of both baths are qubits, with Hamiltonian $H_i = \omega|1\rangle\langle1|$, where $i\in\{A_n, B_n, S\}$. Similar to the previous example, the intra-bath collisions are given by $U_{AA} = e^{igH_{AA}}$ and $U_{BB} = e^{igH_{BB}}$, where $H_{AA} = H^{+-}_{{A_n}{A_{n-1}}}$, and $H_{BB} = H^{+-}_{{B_n}{B_{n-1}}}$ for $B$. Again, the probe-mediated collisions are given by $U_{SI} = e^{i\kappa_I H_{SI}}$ where $H_{SI} = H^{+-}_{SI_{n}}$ with $I = A,B$. 

We initialize all auxiliary units of baths $A$ and $B$ in Gibbs states at inverse temperatures $\beta_A$ and $\beta_B$, respectively. The probe is also initialized in a Gibbs state at inverse temperature $\beta_A$. Although the derivation of the 
{XFT} is independent of the initial probe state, we choose this equilibrium preparation for concreteness. The baths and the probe then evolve according to the collisional dynamics comprising both intra-bath and probe-mediated inter-bath interactions. After sufficiently many collisions, the joint auxiliary units and probe state reach a steady state in which the joint state of the probe and the $n-1$-th auxiliary units is identical to that of the probe and the $n$-th auxiliary units before the subsequent collision step. The {generalized exchange fluctuation theorem} for the probe-mediated protocol is evaluated in this steady-state regime.

\begin{figure}[!h]
    \centering
    \includegraphics[width = 0.45\textwidth]{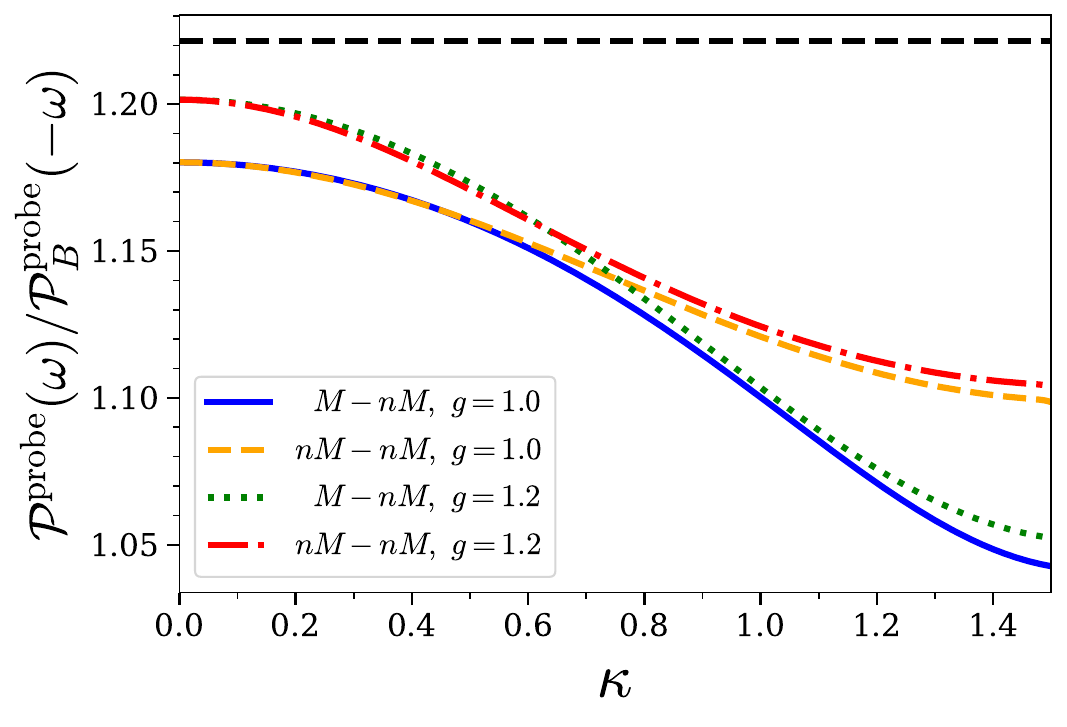}
    \caption{\textbf{XFT for probe-mediated bath-bath interaction.} The left-hand side of Eq.~\eqref{eq:XFT2} is plotted as a function of the intra-bath coupling strength $\kappa$. The bath auxiliaries are qubits with energy splitting $\omega=1.0$, and the probe is also taken to be a qubit with the same energy splitting. The probe is initially prepared in the Gibbs state at inverse temperature $\beta_A$. Results are shown for bath-bath coupling strengths $g=1.0$ and $g=1.2$, comparing a Markovian and a non-Markovian bath $A$, while bath $B$ remains non-Markovian. The inverse temperatures are fixed at $\beta_A=1.0$ and $\beta_B=0.8$. The black dashed line represents the conventional Markovian XFT prediction, $e^{\Delta\beta\omega}$, obtained in the absence of the probe and non-Markovian corrections. {Quantities plotted on both axes are dimensionless.}
}
\label{fig4-XFT2}
\end{figure}

% In the previous subsection, we discussed the evaluation of the BLP measure of non-Markovianity for the baths considered in this work by considering a qubit $Z$. As shown earlier, the degree of non-Markovianity exhibits a periodic dependence on the strength of the intra-bath interaction. 
In the previous subsection, we characterized the non-Markovianity of the bath $B$ using the BLP measure $\mathcal{N}$. To this end, we considered a probe qubit $Z$ undergoing collisional dynamics with the qubit auxiliaries of the bath, where non-Markovian effects arise from intra-bath collisions between successive auxiliaries. As shown in Fig.~\ref{fig2-NM}, the resulting degree of non-Markovianity exhibits a periodic dependence on the intra-bath interaction strength $\kappa$.
We first validate Eq.~\eqref{eq:XFT2} by independently evaluating its left- and right-hand sides and confirming their agreement. Fig.~\ref{fig4-XFT2} shows the ratio $\mP^{\prbe}(\omega)/\mP_B^{\prbe}(-\omega)$ over half a period of the intra-bath interaction, where the BLP measure of non-Markovianity increases monotonically. In this regime, the probability ratio decreases monotonically with increasing non-Markovianity, indicating an enhanced relative probability of reverse heat-transfer events. Consistent with the direct bath-bath interaction, the suppression is more pronounced when only one bath is non-Markovian than when both baths exhibit non-Markovian dynamics.
 
\section{Conclusion}
\label{Sec4}
We have derived 
% \textcolor{orange}{\sout{modified exchange fluctuation theorems}} 
{
% \color{cyan}
{generalized exchange fluctuation theroem}} for non-Markovian thermal baths within a microscopic collisional-model framework. By incorporating intra-bath interactions that generate memory between successive collisions, we obtain exact 
% \textcolor{orange}{\sout{corrections}}
% \SaM
{modification} to the standard Jarzynski-W{\'o}jcik exchange fluctuation theorem for both direct bath-bath and probe-mediated heat exchange. These modifications are determined by factor consisting of the conditional averages of the exponential of stochastic free-energy and stochastic mutual information differences associated with the steady states of the colliding bath auxiliaries and reduce to unity in the Markovian limit, recovering the conventional exchange fluctuation theorem.

% Using \textcolor{orange}{qubit baths} 
Using baths with qubit auxiliary units, as an illustrative example, we showed that increasing bath memory enhances the probability of reverse heat-transfer events, providing a clear thermodynamic signature of non-Markovianity. Our results establish a direct connection between environmental memory and non-equilibrium fluctuation relations within a microscopic framework relevant to quantum reservoirs. The present approach can be extended to transient dynamics, multiple conserved quantities, strong system-bath coupling, and many-body reservoirs, offering a route toward understanding fluctuation relations in correlated quantum environments.

\acknowledgements
S. Mondal acknowledges support from the Infosys scholarship for senior students at Harish-Chandra Research Institute. U. S. acknowledges financial support from the Anusandhan National Research Foundation (ANRF), Government of India, under the Grant No. ANRF/ARG/2025/004617/PS.
\appendix
\section*{Appendix}

\section{Reverse trajectory and its probability}
\label{AppA}

To derive the modified exchange fluctuation theorem, we first define the backward process corresponding to the collisional dynamics. Following the construction of Ref.~\cite{Scandi2026}, the backward dynamics is obtained by time-reversing the initial state and the sequence of collisions. We summarize the construction below.

For the reverse trajectory, the bath auxiliary units in the direct bath-bath interaction case have the initial state 
\begin{align}
    \tilde{\rho}^{\dir}_0 = \Theta(\gamma_{A_n}^{\beta_A}\otimes \gamma_{B_n}^{\beta_B}\otimes\rho^{ss}_{A_{n-1}B_{n-1}})\Theta^{-1}.
\end{align}
Here, $\Theta$ is the anti-unitary time-reversal operator and the states $\gamma_{A_n}^{\beta_A}$ and $\gamma_{B_n}^{\beta_B}$ are the corresponding equilibrium Gibbs state of the $n$-th auxiliary units of the baths $A$ and $B$, respectively. The state $\rho^{ss}_{A_{n-1}B_{n-1}}$ is the joint steady state of the $(n-1)$-th auxiliary units. In the case of the probe-mediated collision, we also have the probe system as well, such that the initial state of the reverse trajectory is given by
\begin{align}
    \tilde{\rho}^{\prbe}_0 = \Theta(\gamma_{A_n}^{\beta_A}\otimes \gamma_{B_n}^{\beta_B}\otimes\rho^{ss}_{SA_{n-1}B_{n-1}})\Theta^{-1}.
\end{align}
Under time-reversal, the Hamiltonian of the two baths transforms as $\Theta H_A \Theta^{-1}$ and $\Theta H_B \Theta^{-1}$. Consequently, for TPM the initial measurement is done using the projectors $\Theta \Pi_A\Theta^{-1}$ and $\Theta \Pi_B\Theta^{-1}$ for both the $n$-th and $(n-1)$-th units of both baths. For the probe-mediated collisions case, the probe is measured using $\Theta \Pi_S\Theta^{-1}$. 
Thus, probability of obtaining $|j_A n_A\xi_B\nu_B\rangle$ in the reverse trajectory after the first energy measurement is $$\tilde{p}^{\dir}(j,n,\xi,\nu) = p_n^{\beta_A}p_\mu^{\beta_B}q_{j\xi}^{AB}$$ for the direct bath-bath interaction. Similarly, for the probe-mediated interaction case, the probability to obtain $|j_An_Al_S\xi_B\nu_B\rangle$ in the reverse trajectory is given by
$$\tilde{p}^{\prbe}(j,n,l,\xi,\nu) = p_n^{\beta_A}p_\mu^{\beta_B}q_{lj\xi}^{SAB}.$$

Now let us consider the unitary evolution involving the collisional model for the backward process.
For this, let us consider the total Hamiltonian leading to collisions in the forward process for the direct bath-bath interaction, to be
\begin{align}
H^{\dir}_{\text{int}}(t)=
\begin{cases}
\tilde{\kappa}_AH_{AA} + \tilde{\kappa}_BH_{BB}, & 0 \le t < t_1,\\
\tilde{g}H_{AB}, & t_1 \le t < t_2.
\end{cases}
\end{align}
Consequently, we have the unitary evolution as
\begin{align}
\mathcal{U} &= \exp\left[{-(i/\hbar)\int_0^{t_2} H_{\text{int}}^{\dir}(t)dt}\right] \nonumber\\
&= e^{-(i/\hbar)\tilde{g}H_{AB}(t_2-t_1)}\left(e^{(-i/\hbar)({\tilde{\kappa}_AH_{AA}}+ {\tilde{\kappa}_BH_{BB})t_1}}\right) \nonumber \\
&= e^{{-ig}H_{AB}} \left(e^{{-i\kappa}_AH_{AA}}\otimes e^{{-i\kappa}_BH_{BB}}\right)\nonumber\\
&= U_{AB}(U_{AA}\otimes U_{BB}).
\end{align}
Here, we introduce $g \coloneqq \tilde{g}(t_2-t_1)/\hbar$ and $\kappa_{A(B)}\coloneqq\tilde\kappa_{A(B)}t_1/\hbar$. 

Since the backward trajectory is defined by reversing the temporal ordering of the interactions, the interaction Hamiltonian satisfies
$\tilde{H}^{\dir}_{\text{int}}(t) = \Theta H_{\text{int}}^{\dir}(t_2-t)\Theta^{-1}$~\cite{Campisi2011}. Following this, we have 
\begin{align}
\tilde{\mathcal{U}} &= \exp\left[(-i/\hbar)\int_0^{t_2}\tilde{H}_{\text{int}}^{\dir} (t)dt\right] 
% =  \Theta\left(e^{{\kappa}_AH_{AA}}\otimes e^{{\kappa}_BH_{BB}}\right)e^{{g}H_{AB}}\Theta^{-1} 
\nonumber \\
&=\Theta(U_{AA}\otimes U_{BB})U_{AB}\Theta^{-1} = \Theta\mathcal{U}^\dagger\Theta^{-1}. 
\end{align}
This is physically due to microreversibility together with the reversed ordering of the interaction sequence.

Similarly, for the probe-mediated collisions case, we have $\tilde{\mathbf{U}} = \Theta\mathbf{U}^\dagger\Theta^{-1}$ as well due to the piecewise nature of the Hamiltonian. 
The unitary operation in the probe-mediated interaction is given by $\mathbf{U} = (U_{SA}\otimes U_{SB})(U_{AA}\otimes U_{BB})$. This is generated using a piecewise constant Hamiltonian,
\begin{align}
    H^{\prbe}_{\text{int}}(t)=
\begin{cases}
\tilde{\kappa}'_AH'_{AA} + \tilde{\kappa}'_BH'_{BB}, & 0 \le t < t'_1,\\
\tilde{g}'H'_{SB}, & t'_1 \le t < t'_1 + \Delta_t\\
\tilde{g}'H'_{SA}, & t'_1 + \Delta_t \le t < t'_2.
\end{cases}
\end{align}
Here, $\Delta_t = (t'_2+t'_1)/2$, $U_{oo} = \exp[(-i/\hbar)\tilde{\kappa}_o't_1'H_{oo}'] = \exp[-i\kappa_o H_{oo}'] $ and $U_{So} = \exp[(-i/\hbar)\tilde{g}'(\frac{t'_2-t'_1}{2}) H_{So}'] = \exp[-ig' H_{So}']$, with $o\in\{A,B\}$.
The backward trajectory evolution operator is given by $ \tilde{\mathbf{U}}$, where $\tilde{\mathbf{U}} = \exp\left[(-i/\hbar)\int_0^{t_2'} \tilde{H}_{\text{int}}^{\prbe}(t')dt'\right] = \Theta\mathbf{U}^\dagger\Theta^{-1}$, as $\tilde{H}_{\text{int}}^{\prbe}(t) = \Theta H^{\prbe}_{\text{int}}(t_2 - t)\Theta^{-1}$.

Therefore, the probabilities of the backward trajectories used in the derivation of the modified exchange fluctuation theorem are
\begin{align*}
     P^{\dir}_B(\tilde{\Gamma}) &= p_n^{\beta_A} p_\nu^{\beta_B} q_{j\xi}^{AB} |\langle\mathbf{i}|\mathcal U^\dagger|\mathbf{f}\rangle|^2,\\
      P^{\prbe}_B(\tilde{\Gamma}) &= p_n^{\beta_A} p_\nu^{\beta_B} q_{lj\xi}^{SAB}|\langle\mathbf{I}|\mathbf U^\dagger|\mathbf{F}\rangle|^2.
\end{align*}
\bibliography{ref}

@article{Akagawa2009,
    author = {Akagawa, Shiho and Hatano, Naomichi},
    title = {The Exchange Fluctuation Theorem in Quantum Mechanics},
    journal = {Prog. Theor. Phys.},
    volume = {121},
    number = {6},
    pages = {1157-1172},
    year = {2009},
    month = {06},
    issn = {0033-068X},
    doi = {10.1143/PTP.121.1157},
    url = {https://doi.org/10.1143/PTP.121.1157}
}

@article{Andrieux_2009,
doi = {10.1088/1367-2630/11/4/043014},
url = {https://doi.org/10.1088/1367-2630/11/4/043014},
year = {2009},
month = {apr},
publisher = {},
volume = {11},
number = {4},
pages = {043014},
author = {Andrieux, D and Gaspard, P and Monnai, T and Tasaki, S},
title = {The fluctuation theorem for currents in open quantum systems},
journal = {New J. Phys.}
}

@article{Denzler2018,
  title = {Heat distribution of a quantum harmonic oscillator},
  author = {Denzler, Tobias and Lutz, Eric},
  journal = {Phys. Rev. E},
  volume = {98},
  issue = {5},
  pages = {052106},
  numpages = {5},
  year = {2018},
  month = {Nov},
  publisher = {American Physical Society},
  doi = {10.1103/PhysRevE.98.052106},
  url = {https://link.aps.org/doi/10.1103/PhysRevE.98.052106}
}

@article{Devi2021,
doi = {10.1088/1742-5468/abdd14},
url = {https://doi.org/10.1088/1742-5468/abdd14},
year = {2021},
month = {feb},
publisher = {IOP Publishing and SISSA},
volume = {2021},
number = {2},
pages = {023209},
author = {Usha Devi, A R and Sudha and Rajagopal, A K and Jayannavar, A M},
title = {Heat exchange and fluctuation in Gaussian thermal states in the quantum realm},
journal = {J. Stat. Mech.}
}

@article{Gomez-Marin2006,
  title = {Heat fluctuations in Brownian transducers},
  author = {Gomez-Marin, A. and Sancho, J. M.},
  journal = {Phys. Rev. E},
  volume = {73},
  issue = {4},
  pages = {045101},
  numpages = {4},
  year = {2006},
  month = {Apr},
  publisher = {American Physical Society},
  doi = {10.1103/PhysRevE.73.045101},
  url = {https://link.aps.org/doi/10.1103/PhysRevE.73.045101}
}

@article{Hasegawa2019,
  title = {Fluctuation Theorem Uncertainty Relation},
  author = {Hasegawa, Yoshihiko and Van Vu, Tan},
  journal = {Phys. Rev. Lett.},
  volume = {123},
  issue = {11},
  pages = {110602},
  numpages = {6},
  year = {2019},
  month = {Sep},
  publisher = {American Physical Society},
  doi = {10.1103/PhysRevLett.123.110602},
  url = {https://link.aps.org/doi/10.1103/PhysRevLett.123.110602}
}

@article{Hernandez-Gomez2025,
  title = {Energy exchange statistics and fluctuation theorem for nonthermal asymptotic states},
  author = {Hern\'andez-G\'omez, Santiago and Poggiali, Francesco and Cappellaro, Paola and Cataliotti, Francesco S. and Trombettoni, Andrea and Fabbri, Nicole and Gherardini, Stefano},
  journal = {Phys. Rev. E},
  volume = {111},
  issue = {1},
  pages = {014139},
  numpages = {13},
  year = {2025},
  month = {Jan},
  publisher = {American Physical Society},
  doi = {10.1103/PhysRevE.111.014139},
  url = {https://link.aps.org/doi/10.1103/PhysRevE.111.014139}
}

@article{Jarzynski2004,
  title = {Classical and Quantum Fluctuation Theorems for Heat Exchange},
  author = {Jarzynski, Christopher and W\'ojcik, Daniel K.},
  journal = {Phys. Rev. Lett.},
  volume = {92},
  issue = {23},
  pages = {230602},
  numpages = {4},
  year = {2004},
  month = {Jun},
  publisher = {American Physical Society},
  doi = {10.1103/PhysRevLett.92.230602},
  url = {https://link.aps.org/doi/10.1103/PhysRevLett.92.230602}
}

@article{Jevtic2015,
  title = {Exchange fluctuation theorem for correlated quantum systems},
  author = {Jevtic, Sania and Rudolph, Terry and Jennings, David and Hirono, Yuji and Nakayama, Shojun and Murao, Mio},
  journal = {Phys. Rev. E},
  volume = {92},
  issue = {4},
  pages = {042113},
  numpages = {12},
  year = {2015},
  month = {Oct},
  publisher = {American Physical Society},
  doi = {10.1103/PhysRevE.92.042113},
  url = {https://link.aps.org/doi/10.1103/PhysRevE.92.042113}
}

@article{Kwon2019,
  title = {Fluctuation Theorems for a Quantum Channel},
  author = {Kwon, Hyukjoon and Kim, M. S.},
  journal = {Phys. Rev. X},
  volume = {9},
  issue = {3},
  pages = {031029},
  numpages = {26},
  year = {2019},
  month = {Aug},
  publisher = {American Physical Society},
  doi = {10.1103/PhysRevX.9.031029},
  url = {https://link.aps.org/doi/10.1103/PhysRevX.9.031029}
}

@article{Levy2020,
  title = {Quasiprobability Distribution for Heat Fluctuations in the Quantum Regime},
  author = {Levy, Amikam and Lostaglio, Matteo},
  journal = {PRX Quantum},
  volume = {1},
  issue = {1},
  pages = {010309},
  numpages = {19},
  year = {2020},
  month = {Sep},
  publisher = {American Physical Society},
  doi = {10.1103/PRXQuantum.1.010309},
  url = {https://link.aps.org/doi/10.1103/PRXQuantum.1.010309}
}

@article{Landi2024,
  title = {Current Fluctuations in Open Quantum Systems: Bridging the Gap Between Quantum Continuous Measurements and Full Counting Statistics},
  author = {Landi, Gabriel T. and Kewming, Michael J. and Mitchison, Mark T. and Potts, Patrick P.},
  journal = {PRX Quantum},
  volume = {5},
  issue = {2},
  pages = {020201},
  numpages = {86},
  year = {2024},
  month = {Apr},
  publisher = {American Physical Society},
  doi = {10.1103/PRXQuantum.5.020201},
  url = {https://link.aps.org/doi/10.1103/PRXQuantum.5.020201}
}

@article{Li2025,
  title = {Full-counting statistics and quantum information of dispersive readout with a squeezed environment},
  url = {https://arxiv.org/abs/2512.02531},
  author = {Li,  Ming and Luo,  JunYan and Platero,  Gloria and Engelhardt,  Georg},
  journal = {arXiv:2512.02531},
  year = {2025}
}

@article{Manzano2016,
  title = {Entropy production and thermodynamic power of the squeezed thermal reservoir},
  author = {Manzano, Gonzalo and Galve, Fernando and Zambrini, Roberta and Parrondo, Juan M. R.},
  journal = {Phys. Rev. E},
  volume = {93},
  issue = {5},
  pages = {052120},
  numpages = {10},
  year = {2016},
  month = {May},
  publisher = {American Physical Society},
  doi = {10.1103/PhysRevE.93.052120},
  url = {https://link.aps.org/doi/10.1103/PhysRevE.93.052120}
}

@article{Nicolin2011,
  title = {Quantum fluctuation theorem for heat exchange in the strong coupling regime},
  author = {Nicolin, Lena and Segal, Dvira},
  journal = {Phys. Rev. B},
  volume = {84},
  issue = {16},
  pages = {161414},
  numpages = {4},
  year = {2011},
  month = {Oct},
  publisher = {American Physical Society},
  doi = {10.1103/PhysRevB.84.161414},
  url = {https://link.aps.org/doi/10.1103/PhysRevB.84.161414}
}

@article{Rodrigues2025,
  title={Far-from-equilibrium thermodynamics of non-Abelian thermal states},
  author={Rodrigues, Franklin LS and Lutz, Eric},
  journal={arXiv:2510.04788},
  url = {https://arxiv.org/abs/2510.04788},
  year={2025}
}

@article{Sarmah2023,
  title={Nonequilibrium fluctuations in boson transport through squeezed reservoirs},
  author={Sarmah, Manash Jyoti and Bansal, Akanksha and Goswami, Himangshu Prabal},
  journal={Physica A},
  volume={615},
  pages={128620},
  year={2023},
  url = {https://doi.org/10.1016/j.physa.2023.128620},
  publisher={Elsevier}
}

@article{Scandi2026,
  title = {Universal Statistics of Charge Exchanges in Non-Abelian Quantum Transport},
  author = {Scandi, Matteo and Manzano, Gonzalo},
  journal = {Phys. Rev. Lett.},
  volume = {136},
  issue = {15},
  pages = {150403},
  numpages = {7},
  year = {2026},
  month = {Apr},
  publisher = {American Physical Society},
  doi = {10.1103/n2gp-8bx9},
  url = {https://link.aps.org/doi/10.1103/n2gp-8bx9}
}

@article{Sone2023,
    author = {Sone, Akira and Soares-Pinto, Diogo O. and Deffner, Sebastian},
    title = {Exchange fluctuation theorems for strongly interacting quantum pumps},
    journal = {AVS Quantum Sci.},
    volume = {5},
    number = {3},
    pages = {032001},
    year = {2023},
    month = {07},
    issn = {2639-0213},
    doi = {10.1116/5.0152186},
    url = {https://doi.org/10.1116/5.0152186}
}

@article{Timpanaro2019,
  title = {Thermodynamic Uncertainty Relations from Exchange Fluctuation Theorems},
  author = {Timpanaro, Andr\'e M. and Guarnieri, Giacomo and Goold, John and Landi, Gabriel T.},
  journal = {Phys. Rev. Lett.},
  volume = {123},
  issue = {9},
  pages = {090604},
  numpages = {6},
  year = {2019},
  month = {Aug},
  publisher = {American Physical Society},
  doi = {10.1103/PhysRevLett.123.090604},
  url = {https://link.aps.org/doi/10.1103/PhysRevLett.123.090604}
}

@article{Upadhyaya2024,
  title = {Non-Abelian Transport Distinguishes Three Usually Equivalent Notions of Entropy Production},
  author = {Upadhyaya, Twesh and Braasch, William F. and Landi, Gabriel T. and Halpern, Nicole Yunger},
  journal = {PRX Quantum},
  volume = {5},
  issue = {3},
  pages = {030355},
  numpages = {25},
  year = {2024},
  month = {Sep},
  publisher = {American Physical Society},
  doi = {10.1103/PRXQuantum.5.030355},
  url = {https://link.aps.org/doi/10.1103/PRXQuantum.5.030355}
}

@article{Wei2018,
  title = {Fluctuation relations for heat exchange in the generalized Gibbs ensemble},
  volume = {13},
  ISSN = {2095-0470},
  url = {http://dx.doi.org/10.1007/s11467-018-0822-y},
  number = {5},
  journal = {Front. Phys.},
  publisher = {China Engineering Science Press Co. Ltd.},
  author = {Wei,  Bo-Bo},
  pages={130510},
  year = {2018},
  month = sep 
}

@article{Wu2024,
  title = {Generalized Quantum Fluctuation Theorem for Energy Exchange},
  author = {Wu, Wei and An, Jun-Hong},
  journal = {Phys. Rev. Lett.},
  volume = {133},
  issue = {5},
  pages = {050401},
  numpages = {7},
  year = {2024},
  month = {Jul},
  publisher = {American Physical Society},
  doi = {10.1103/PhysRevLett.133.050401},
  url = {https://link.aps.org/doi/10.1103/PhysRevLett.133.050401}
}

@article{PhysRevE.103.042143,
  title = {Statistical properties of the heat flux between two nonequilibrium steady-state thermostats},
  author = {Lameche, Mona and Naert, Antoine},
  journal = {Phys. Rev. E},
  volume = {103},
  issue = {4},
  pages = {042143},
  numpages = {6},
  year = {2021},
  month = {Apr},
  publisher = {American Physical Society},
  doi = {10.1103/PhysRevE.103.042143},
  url = {https://link.aps.org/doi/10.1103/PhysRevE.103.042143}
}

@article{Yadalam2022,
  title = {Counting statistics of energy transport across squeezed thermal reservoirs},
  author = {Yadalam, Hari Kumar and Agarwalla, Bijay Kumar and Harbola, Upendra},
  journal = {Phys. Rev. A},
  volume = {105},
  issue = {6},
  pages = {062219},
  numpages = {14},
  year = {2022},
  month = {Jun},
  publisher = {American Physical Society},
  doi = {10.1103/PhysRevA.105.062219},
  url = {https://link.aps.org/doi/10.1103/PhysRevA.105.062219}
}

@article{Campisi2011,
  title = {Colloquium: Quantum fluctuation relations: Foundations and applications},
  author = {Campisi, Michele and H\"anggi, Peter and Talkner, Peter},
  journal = {Rev. Mod. Phys.},
  volume = {83},
  issue = {3},
  pages = {771--791},
  numpages = {0},
  year = {2011},
  month = {Jul},
  publisher = {American Physical Society},
  doi = {10.1103/RevModPhys.83.771},
  url = {https://link.aps.org/doi/10.1103/RevModPhys.83.771}
}

@article{Esposito2009,
  title = {Nonequilibrium fluctuations, fluctuation theorems, and counting statistics in quantum systems},
  author = {Esposito, Massimiliano and Harbola, Upendra and Mukamel, Shaul},
  journal = {Rev. Mod. Phys.},
  volume = {81},
  issue = {4},
  pages = {1665--1702},
  numpages = {0},
  year = {2009},
  month = {Dec},
  publisher = {American Physical Society},
  doi = {10.1103/RevModPhys.81.1665},
  url = {https://link.aps.org/doi/10.1103/RevModPhys.81.1665}
}

@article{Manzano2022,
  title = {Quantum thermodynamics under continuous monitoring: A general framework},
  volume = {4},
  ISSN = {2639-0213},
  url = {http://dx.doi.org/10.1116/5.0079886},
  number = {2},
  journal = {AVS Quantum Sci.},
  pages = {025302},
  publisher = {American Vacuum Society},
  author = {Manzano,  Gonzalo and Zambrini,  Roberta},
  year = {2022},
  month = may 
}

@article{Salazar:2025qfs,
    author = {Salazar, Domingos S. P.},
    title = {Matrix Thermodynamic Uncertainty Relation for Non-Abelian Charge Transport},
    journal = {arXiv:2512.24956},
    url = {https://doi.org/10.48550/arXiv.2512.24956},
    year = {2025}
}

@article{gy6n-5x26,
  title = {Anomalous flow in correlated quantum systems: No-go result and multiple-charge scenario},
  author = {Guan, Rui and Liu, Junjie},
  journal = {Phys. Rev. A},
  volume = {112},
  issue = {2},
  pages = {022220},
  numpages = {15},
  year = {2025},
  month = {Aug},
  publisher = {American Physical Society},
  doi = {10.1103/gy6n-5x26},
  url = {https://link.aps.org/doi/10.1103/gy6n-5x26}
}

@article{PhysRevE.108.054109,
  title = {Exploring quasiprobability approaches to quantum work in the presence of initial coherence: Advantages of the Margenau-Hill distribution},
  author = {Pei, Ji-Hui and Chen, Jin-Fu and Quan, H. T.},
  journal = {Phys. Rev. E},
  volume = {108},
  issue = {5},
  pages = {054109},
  numpages = {14},
  year = {2023},
  month = {Nov},
  publisher = {American Physical Society},
  doi = {10.1103/PhysRevE.108.054109},
  url = {https://link.aps.org/doi/10.1103/PhysRevE.108.054109}
}

@article{McElvogue2026,
  title = {Nonequilibrium steady states in multibath quantum collision models},
  author = {McElvogue, Ronan and Mitchell, Andrew K. and Landi, Gabriel T. and Campbell, Steve},
  journal = {Phys. Rev. A},
  volume = {113},
  issue = {2},
  pages = {022206},
  numpages = {12},
  year = {2026},
  month = {Feb},
  publisher = {American Physical Society},
  doi = {10.1103/ypyy-ql4c},
  url = {https://link.aps.org/doi/10.1103/ypyy-ql4c}
}

@misc{Wolf2012,
  author       = {Michael M. Wolf},
  title        = {Quantum Channels \& Operations: Guided Tour},
  year         = {2012},
  note         = {Lecture notes},
 url = {https://mediatum.ub.tum.de/doc/1701036/1701036.pdf}
}

@article{Scarani2002,
  title = {Thermalizing Quantum Machines: Dissipation and Entanglement},
  volume = {88},
  ISSN = {1079-7114},
  url = {http://dx.doi.org/10.1103/PhysRevLett.88.097905},
  number = {9},
  journal = {Phys. Rev. Lett.},
  publisher = {American Physical Society (APS)},
  author = {Scarani,  Valerio and Ziman,  Mário and Štelmachovič,  Peter and Gisin,  Nicolas and Bužek,  Vladimír},
  year = {2002},
  month = Feb 
}

@article{Saha2024,
  title = {Quantum homogenization in non-Markovian collisional model},
  volume = {26},
  ISSN = {1367-2630},
  url = {http://dx.doi.org/10.1088/1367-2630/ad212f},
  DOI = {10.1088/1367-2630/ad212f},
  number = {2},
  journal = {New J. Phys.},
  publisher = {IOP Publishing},
  author = {Saha,  Tanmay and Das,  Arpan and Ghosh,  Sibasish},
  year = {2024},
  month = Feb,
  pages = {023011}
}

@article{Ziman2002,
  title = {Diluting quantum information: An analysis of information transfer in system-reservoir interactions},
  volume = {65},
  ISSN = {1094-1622},
  url = {http://dx.doi.org/10.1103/PhysRevA.65.042105},
  number = {4},
  journal = {Phys. Rev. A},
  publisher = {American Physical Society (APS)},
  author = {Ziman,  M. and Štelmachovič,  P. and Bužek,  V. and Hillery,  M. and Scarani,  V. and Gisin,  N.},
  year = {2002},
  month = Mar 
}

@article{Ciccarello2022,
title = {Quantum collision models: Open system dynamics from repeated interactions},
journal = {Phys. Rep.},
volume = {954},
pages = {1-70},
year = {2022},
url = {https://www.sciencedirect.com/science/article/pii/S0370157322000035},
author = {Francesco Ciccarello and Salvatore Lorenzo and Vittorio Giovannetti and G. Massimo Palma}
}

@article{Breuer2009,
  title = {Measure for the Degree of Non-Markovian Behavior of Quantum Processes in Open Systems},
  author = {Breuer, Heinz-Peter and Laine, Elsi-Mari and Piilo, Jyrki},
  journal = {Phys. Rev. Lett.},
  volume = {103},
  issue = {21},
  pages = {210401},
  numpages = {4},
  year = {2009},
  month = {Nov},
  publisher = {American Physical Society},
  doi = {10.1103/PhysRevLett.103.210401},
  url = {https://link.aps.org/doi/10.1103/PhysRevLett.103.210401}
}

@article{Laine2010,
  title = {Measure for the non-Markovianity of quantum processes},
  author = {Laine, Elsi-Mari and Piilo, Jyrki and Breuer, Heinz-Peter},
  journal = {Phys. Rev. A},
  volume = {81},
  issue = {6},
  pages = {062115},
  numpages = {8},
  year = {2010},
  month = {Jun},
  publisher = {American Physical Society},
  doi = {10.1103/PhysRevA.81.062115},
  url = {https://link.aps.org/doi/10.1103/PhysRevA.81.062115}
}

@article{McCloskey2014,
  title = {Non-Markovianity and system-environment correlations in a microscopic collision model},
  author = {McCloskey, Ruari and Paternostro, Mauro},
  journal = {Phys. Rev. A},
  volume = {89},
  issue = {5},
  pages = {052120},
  numpages = {6},
  year = {2014},
  month = {May},
  publisher = {American Physical Society},
  doi = {10.1103/PhysRevA.89.052120},
  url = {https://link.aps.org/doi/10.1103/PhysRevA.89.052120}
}

@Article{Senyasa2022,
author = {{\c{S}}enya{\c{s}}a, H. T. and Kesgin, {\c{S}}. and Karpat, G. and {\c{C}}akmak, B.},
TITLE = {Entropy Production in Non-Markovian Collision Models: Information Backflow vs. System-Environment Correlations},
JOURNAL = {Entropy},
VOLUME = {24},
YEAR = {2022},
NUMBER = {6},
ARTICLE-NUMBER = {824},
URL = {https://www.mdpi.com/1099-4300/24/6/824}
}

@article{Manzano2022PRXQ,
  title = {Non-Abelian Quantum Transport and Thermosqueezing Effects},
  author = {Manzano, Gonzalo and Parrondo, Juan M.R. and Landi, Gabriel T.},
  journal = {PRX Quantum},
  volume = {3},
  issue = {1},
  pages = {010304},
  numpages = {14},
  year = {2022},
  month = {Jan},
  publisher = {American Physical Society},
  doi = {10.1103/PRXQuantum.3.010304},
  url = {https://link.aps.org/doi/10.1103/PRXQuantum.3.010304}
}

@article{Lacroix2025,
  title = {Making quantum collision models exact},
  volume = {8},
  ISSN = {2399-3650},
  url = {http://dx.doi.org/10.1038/s42005-025-02201-2},
  number = {1},
  journal = {Commun. Phys.},
  pages = {268},
  publisher = {Springer Science and Business Media LLC},
  author = {Lacroix,  Thibaut and Cilluffo,  Dario and Huelga,  Susana F. and Plenio,  Martin B.},
  year = {2025}
}

@book{Deffnerbook,
author = {Deffner, Sebastian and Campbell, Steve},
title = {Quantum Thermodynamics},
publisher = {Morgan \& Claypool Publishers},
year = {2019},
series = {2053-2571},
isbn = {978-1-64327-658-8},
url = {https://doi.org/10.1088/2053-2571/ab21c6},
doi = {10.1088/2053-2571/ab21c6}
}

@book{Gemmer2004,
  title = {Quantum Thermodynamics: Emergence of Thermodynamic Behavior Within Composite Quantum Systems},
  ISBN = {9783540445135},
  ISSN = {1616-6361},
  url = {http://dx.doi.org/10.1007/b98082},
  DOI = {10.1007/b98082},
  journal = {Lecture Notes in Physics},
  publisher = {Springer Berlin Heidelberg},
  author = {Gemmer,  J. and Michel,  M. and Mahler,  G.},
  year = {2004}
}

@book{Strasbergbook,
    author = {Strasberg, Philipp},
    title = {Quantum Stochastic Thermodynamics: Foundations and Selected Applications},
    publisher = {Oxford University Press},
    year = {2022},
    month = {01},
    isbn = {9780192895585},
    doi = {10.1093/oso/9780192895585.001.0001},
    url = {https://doi.org/10.1093/oso/9780192895585.001.0001}
}

@book{Seifertbook, 
place={Cambridge}, 
title={Stochastic Thermodynamics}, 
publisher={Cambridge University Press}, 
author={Seifert, Udo}, 
year={2025}
}

@book{Binderbook,
  title = {Thermodynamics in the Quantum Regime: Fundamental Aspects and New Directions},
  ISBN = {9783319990460},
  ISSN = {2365-6425},
  url = {http://dx.doi.org/10.1007/978-3-319-99046-0},
  DOI = {10.1007/978-3-319-99046-0},
  journal = {Fundamental Theories of Physics},
  author = {},
  editor = {},
  publisher = {Springer International Publishing},
  year = {2018}
}

@article{BatteriesRMP,
  title = {Colloquium: Quantum batteries},
  author = {Campaioli, Francesco and Gherardini, Stefano and Quach, James Q. and Polini, Marco and Andolina, Gian Marcello},
  journal = {Rev. Mod. Phys.},
  volume = {96},
  issue = {3},
  pages = {031001},
  numpages = {30},
  year = {2024},
  month = {Jul},
  publisher = {American Physical Society},
  doi = {10.1103/RevModPhys.96.031001},
  url = {https://link.aps.org/doi/10.1103/RevModPhys.96.031001}
}

@article{Cangemi2024,
title = {Quantum engines and refrigerators},
journal = {Phys. Rep.},
volume = {1087},
pages = {1-71},
year = {2024},
issn = {0370-1573},
doi = {https://doi.org/10.1016/j.physrep.2024.07.001},
author = {Loris Maria Cangemi and Chitrak Bhadra and Amikam Levy}
}

@article{Linden2010,
  title = {How Small Can Thermal Machines Be? The Smallest Possible Refrigerator},
  author = {Linden, Noah and Popescu, Sandu and Skrzypczyk, Paul},
  journal = {Phys. Rev. Lett.},
  volume = {105},
  issue = {13},
  pages = {130401},
  numpages = {4},
  year = {2010},
  month = {Sep},
  publisher = {American Physical Society},
  doi = {10.1103/PhysRevLett.105.130401},
  url = {https://link.aps.org/doi/10.1103/PhysRevLett.105.130401}
}

@article{Alicki1979,
url = {https://doi.org/10.1088/0305-4470/12/5/007},
year = {1979},
month = {may},
publisher = {},
volume = {12},
number = {5},
pages = {L103},
author = {R Alicki},
title = {The quantum open system as a model of the heat engine},
journal = {J. Phys. A: Math. Gen}
}

@article{Kosloff1984,
  title = {A quantum mechanical open system as a model of a heat engine},
  volume = {80},
  ISSN = {1089-7690},
  url = {http://dx.doi.org/10.1063/1.446862},
  DOI = {10.1063/1.446862},
  number = {4},
  journal = {J. Chem. Phys},
  publisher = {AIP Publishing},
  author = {Kosloff,  Ronnie},
  year = {1984},
  month = Feb,
  pages = {1625–1631}
}

@article{Quan2007,
  title = {Quantum thermodynamic cycles and quantum heat engines},
  author = {Quan, H. T. and Liu, Yu-xi and Sun, C. P. and Nori, Franco},
  journal = {Phys. Rev. E},
  volume = {76},
  issue = {3},
  pages = {031105},
  numpages = {18},
  year = {2007},
  month = {Sep},
  publisher = {American Physical Society},
  doi = {10.1103/PhysRevE.76.031105},
  url = {https://link.aps.org/doi/10.1103/PhysRevE.76.031105}
}

@article{Alicki2013,
  title = {Entanglement boost for extractable work from ensembles of quantum batteries},
  author = {Alicki, Robert and Fannes, Mark},
  journal = {Phys. Rev. E},
  volume = {87},
  issue = {4},
  pages = {042123},
  numpages = {4},
  year = {2013},
  month = {Apr},
  publisher = {American Physical Society},
  doi = {10.1103/PhysRevE.87.042123},
  url = {https://link.aps.org/doi/10.1103/PhysRevE.87.042123}
}

@article{Mitchison2019,
  title = {Quantum thermal absorption machines: refrigerators,  engines and clocks},
  volume = {60},
  ISSN = {1366-5812},
  url = {http://dx.doi.org/10.1080/00107514.2019.1631555},
  DOI = {10.1080/00107514.2019.1631555},
  number = {2},
  journal = {Contemporary Physics},
  publisher = {Informa UK Limited},
  author = {Mitchison,  Mark T.},
  year = {2019},
  month = Apr,
  pages = {164}
}

@article{Skrzypczyk2011,
  title = {The smallest refrigerators can reach maximal efficiency},
  volume = {44},
  ISSN = {1751-8121},
  url = {http://dx.doi.org/10.1088/1751-8113/44/49/492002},
  DOI = {10.1088/1751-8113/44/49/492002},
  number = {49},
  journal = {J. Phys. A: Math. Theor},
  publisher = {IOP Publishing},
  author = {Skrzypczyk,  Paul and Brunner,  Nicolas and Linden,  Noah and Popescu,  Sandu},
  year = {2011},
  month = Nov,
  pages = {492002}
}

@article{Levy2012,
  title = {Quantum Absorption Refrigerator},
  author = {Levy, Amikam and Kosloff, Ronnie},
  journal = {Phys. Rev. Lett.},
  volume = {108},
  issue = {7},
  pages = {070604},
  numpages = {5},
  year = {2012},
  month = {Feb},
  publisher = {American Physical Society},
  doi = {10.1103/PhysRevLett.108.070604},
  url = {https://link.aps.org/doi/10.1103/PhysRevLett.108.070604}
}

@article{Jarzynski1997,
  title = {Nonequilibrium Equality for Free Energy Differences},
  author = {Jarzynski, C.},
  journal = {Phys. Rev. Lett.},
  volume = {78},
  issue = {14},
  pages = {2690--2693},
  numpages = {0},
  year = {1997},
  month = {Apr},
  publisher = {American Physical Society},
  doi = {10.1103/PhysRevLett.78.2690},
  url = {https://link.aps.org/doi/10.1103/PhysRevLett.78.2690}
}

@article{Crooks1999,
  title = {Entropy production fluctuation theorem and the nonequilibrium work relation for free energy differences},
  author = {Crooks, Gavin E.},
  journal = {Phys. Rev. E},
  volume = {60},
  issue = {3},
  pages = {2721--2726},
  numpages = {0},
  year = {1999},
  month = {Sep},
  publisher = {American Physical Society},
  doi = {10.1103/PhysRevE.60.2721},
  url = {https://link.aps.org/doi/10.1103/PhysRevE.60.2721}
}

@article{Evans2002,
  title = {The Fluctuation Theorem},
  volume = {51},
  ISSN = {1460-6976},
  url = {http://dx.doi.org/10.1080/00018730210155133},
  DOI = {10.1080/00018730210155133},
  number = {7},
  journal = {Advances in Physics},
  publisher = {Informa UK Limited},
  author = {Evans,  Denis J. and Searles,  Debra J.},
  year = {2002},
  month = Nov,
  pages = {1529}
}

@article{Seifert_2012,
doi = {10.1088/0034-4885/75/12/126001},
url = {https://doi.org/10.1088/0034-4885/75/12/126001},
year = {2012},
month = {nov},
publisher = {IOP Publishing},
volume = {75},
number = {12},
pages = {126001},
author = {Seifert, Udo},
title = {Stochastic thermodynamics, fluctuation theorems and molecular machines},
journal = {Rep. Prog. Phys.}
}

@article{kurchan2000,
  title={A quantum fluctuation theorem},
  author={Kurchan, Jorge},
  journal={arXiv cond-mat/0007360},
  year={2000},
  url = {https://doi.org/10.48550/arXiv.cond-mat/0007360}
}

@article{tasaki2000,
  title={Jarzynski relations for quantum systems and some applications},
  author={Tasaki, Hal},
  journal={arXiv preprint cond-mat/0009244},
  year={2000}
}

@article{Talkner2007,
  title = {Fluctuation theorems: Work is not an observable},
  author = {Talkner, Peter and Lutz, Eric and H\"anggi, Peter},
  journal = {Phys. Rev. E},
  volume = {75},
  issue = {5},
  pages = {050102(R)},
  numpages = {2},
  year = {2007},
  month = {May},
  publisher = {American Physical Society},
  doi = {10.1103/PhysRevE.75.050102},
  url = {https://link.aps.org/doi/10.1103/PhysRevE.75.050102}
}

@article{Talkner_2007JPA,
doi = {10.1088/1751-8113/40/26/F08},
url = {https://doi.org/10.1088/1751-8113/40/26/F08},
year = {2007},
month = {jun},
publisher = {},
volume = {40},
number = {26},
pages = {F569},
author = {Talkner, Peter and Hänggi, Peter},
title = {The Tasaki–Crooks quantum fluctuation theorem},
journal = {J. Phys. A: Math. Theor.}
}

@article{Micadei2020,
  title = {Quantum Fluctuation Theorems beyond Two-Point Measurements},
  author = {Micadei, Kaonan and Landi, Gabriel T. and Lutz, Eric},
  journal = {Phys. Rev. Lett.},
  volume = {124},
  issue = {9},
  pages = {090602},
  numpages = {6},
  year = {2020},
  month = {Mar},
  publisher = {American Physical Society},
  doi = {10.1103/PhysRevLett.124.090602},
  url = {https://link.aps.org/doi/10.1103/PhysRevLett.124.090602}
}

@article{Micadei2021,
  title = {Experimental Validation of Fully Quantum Fluctuation Theorems Using Dynamic Bayesian Networks},
  author = {Micadei, Kaonan and Peterson, John P. S. and Souza, Alexandre M. and Sarthour, Roberto S. and Oliveira, Ivan S. and Landi, Gabriel T. and Serra, Roberto M. and Lutz, Eric},
  journal = {Phys. Rev. Lett.},
  volume = {127},
  issue = {18},
  pages = {180603},
  numpages = {6},
  year = {2021},
  month = {Oct},
  publisher = {American Physical Society},
  doi = {10.1103/PhysRevLett.127.180603},
  url = {https://link.aps.org/doi/10.1103/PhysRevLett.127.180603}
}

@article{Gherardini2021,
  title = {End-point measurement approach to assess quantum coherence in energy fluctuations},
  author = {Gherardini, S. and Belenchia, A. and Paternostro, M. and Trombettoni, A.},
  journal = {Phys. Rev. A},
  volume = {104},
  issue = {5},
  pages = {L050203},
  numpages = {5},
  year = {2021},
  month = {Nov},
  publisher = {American Physical Society},
  doi = {10.1103/PhysRevA.104.L050203},
  url = {https://link.aps.org/doi/10.1103/PhysRevA.104.L050203}
}

@article{HernndezGmez2023,
  title = {Experimental signature of initial quantum coherence on entropy production},
  volume = {9},
  ISSN = {2056-6387},
  url = {http://dx.doi.org/10.1038/s41534-023-00738-0},
  number = {1},
  journal = {npj Quantum Inf.},
  publisher = {Springer Science and Business Media LLC},
  pages = {86},
  author = {Hernández-Gómez,  Santiago and Gherardini,  Stefano and Belenchia,  Alessio and Trombettoni,  Andrea and Paternostro,  Mauro and Fabbri,  Nicole},
  year = {2023}
}

@article{Gianani2023,
  title = {Diagnostics of quantum-gate coherences deteriorated by unitary errors via end-point-measurement statistics},
  volume = {8},
  ISSN = {2058-9565},
  url = {http://dx.doi.org/10.1088/2058-9565/acedca},
  DOI = {10.1088/2058-9565/acedca},
  number = {4},
  journal = {Quantum Sci. Technol.},
  publisher = {IOP Publishing},
  author = {Gianani,  Ilaria and Belenchia,  Alessio and Gherardini,  Stefano and Berardi,  Vincenzo and Barbieri,  Marco and Paternostro,  Mauro},
  year = {2023},
  month = Aug,
  pages = {045018}
}

@article{Artini2026,
  title = {Coherent heat exchange in a prethermalizing open quantum system},
  author = {Artini, Simone and Paternostro, Mauro and Lorenzo, Salvatore},
  journal = {Phys. Rev. A},
  volume = {113},
  issue = {5},
  pages = {052203},
  numpages = {8},
  year = {2026},
  month = {May},
  publisher = {American Physical Society},
  doi = {10.1103/q5vb-dfgj},
  url = {https://link.aps.org/doi/10.1103/q5vb-dfgj}
}

@article{Mondkar2025,
  title={Resource-resolved quantum fluctuation theorems in end-point measurement scheme},
  author={Mondkar, Sukrut and Mondal, Sayan and Sen, Ujjwal},
  journal={arXiv:2512.15928},
  year={2025},
  url = {https://arxiv.org/abs/2512.15928}
}

@article{Lostaglio2018,
  title = {Quantum Fluctuation Theorems, Contextuality, and Work Quasiprobabilities},
  author = {Lostaglio, Matteo},
  journal = {Phys. Rev. Lett.},
  volume = {120},
  issue = {4},
  pages = {040602},
  numpages = {6},
  year = {2018},
  month = {Jan},
  publisher = {American Physical Society},
  doi = {10.1103/PhysRevLett.120.040602},
  url = {https://link.aps.org/doi/10.1103/PhysRevLett.120.040602}
}

@article{Jae2026,
  title = {Operational quasiprobability in quantum thermodynamics: Work extraction by coherence and nonjoint measurability},
  author = {Jae, Jeongwoo and Ryu, Junghee and Ryu, Hoon},
  journal = {Phys. Rev. Res.},
  volume = {8},
  issue = {1},
  pages = {013232},
  numpages = {11},
  year = {2026},
  month = {Mar},
  publisher = {American Physical Society},
  doi = {10.1103/qbs6-nj3l},
  url = {https://link.aps.org/doi/10.1103/qbs6-nj3l}
}

@article{Li2025SciAdv,
    title = {Experimental demonstration of generalized quantum fluctuation theorems in the presence of coherence},
  volume = {11},
  ISSN = {2375-2548},
  url = {http://dx.doi.org/10.1126/sciadv.adq6014},
  number = {22},
  journal = {Sci. Adv.},
  pages = {eadq6014},
  publisher = {American Association for the Advancement of Science (AAAS)},
  author = {Li,  Hui and Xie,  Jie and Kwon,  Hyukjoon and Zhao,  Yixin and Kim,  M. S. and Zhang,  Lijian},
  year = {2025}
}

@article{Ciccarello2013,
  title = {Collision-model-based approach to non-Markovian quantum dynamics},
  author = {Ciccarello, F. and Palma, G. M. and Giovannetti, V.},
  journal = {Phys. Rev. A},
  volume = {87},
  issue = {4},
  pages = {040103(R)},
  numpages = {5},
  year = {2013},
  month = {Apr},
  publisher = {American Physical Society},
  doi = {10.1103/PhysRevA.87.040103},
  url = {https://link.aps.org/doi/10.1103/PhysRevA.87.040103}
}

@article{Santos2020,
  title = {Joint Fluctuation Theorems for Sequential Heat Exchange},
  volume = {22},
  ISSN = {1099-4300},
  url = {http://dx.doi.org/10.3390/e22070763},
  DOI = {10.3390/e22070763},
  number = {7},
  journal = {Entropy},
  publisher = {MDPI AG},
  author = {Santos,  Jader and Timpanaro,  André and Landi,  Gabriel},
  year = {2020},
  pages = {763}
}

@article{Schmidt2023,
  title = {Stochastic Thermodynamics of a Finite Quantum System Coupled to Two Heat Baths},
  volume = {25},
  ISSN = {1099-4300},
  url = {http://dx.doi.org/10.3390/e25030504},
  DOI = {10.3390/e25030504},
  number = {3},
  journal = {Entropy},
  publisher = {MDPI AG},
  author = {Schmidt,  Heinz-J\"{u}rgen and Gemmer,  Jochen},
  year = {2023},
  pages = {504}
}

@article{Esposito2010,
  title = {Entropy production as correlation between system and reservoir},
  volume = {12},
  ISSN = {1367-2630},
  url = {http://dx.doi.org/10.1088/1367-2630/12/1/013013},
  DOI = {10.1088/1367-2630/12/1/013013},
  number = {1},
  journal = {New J. Phys.},
  publisher = {IOP Publishing},
  author = {Esposito,  Massimiliano and Lindenberg,  Katja and Van den Broeck,  Christian},
  year = {2010},
  pages = {013013}
}

@article{PhysRevE.99.012120,
  title = {Non-Markovianity and negative entropy production rates},
  author = {Strasberg, Philipp and Esposito, Massimiliano},
  journal = {Phys. Rev. E},
  volume = {99},
  issue = {1},
  pages = {012120},
  numpages = {24},
  year = {2019},
  month = {Jan},
  publisher = {American Physical Society},
  doi = {10.1103/PhysRevE.99.012120},
  url = {https://link.aps.org/doi/10.1103/PhysRevE.99.012120}
}

@article{mondal2023modified,
  title={Modified Landauer's principle: How much can the Maxwell's demon gain by using general system-environment quantum state?},
  author={Mondal, Sayan and Bhattacharyya, Aparajita and Ghoshal, Ahana and Sen, Ujjwal},
  journal={arXiv:2309.09678},
url = {https://arxiv.org/abs/2309.09678},
  year={2023}
}

@article{jiang2018improved,
  title={Improved Landauer's principle and generalized second law of thermodynamics with initial correlations and non-equilibrium surrounding environments},
  author={Jiang, Ke-Xia and Li, Yuan-Mou and Fan, Heng},
  journal={arXiv:1803.06125},
  url = {https://arxiv.org/abs/1803.06125},
  year={2018}
}

@article{Bera2017,
  title = {Generalized laws of thermodynamics in the presence of correlations},
  volume = {8},
  ISSN = {2041-1723},
  url = {http://dx.doi.org/10.1038/s41467-017-02370-x},
  number = {1},
  journal = {Nat. Commun.},
  pages={2180},
  publisher = {Springer Science and Business Media LLC},
  author = {Bera,  Manabendra N. and Riera,  Arnau and Lewenstein,  Maciej and Winter,  Andreas},
  year = {2017}
}

@article{Aimet2025,
  title = {Experimentally probing Landauer’s principle in the quantum many-body regime},
  volume = {21},
  url = {http://dx.doi.org/10.1038/s41567-025-02930-9},
  DOI = {10.1038/s41567-025-02930-9},
  number = {8},
  journal = {Nat. Phys.},
  publisher = {Springer Science and Business Media LLC},
  author = {Aimet,  Stefan and Tajik,  Mohammadamin and Tournaire,  Gabrielle and Sch\"{u}ttelkopf,  Philipp and Sabino,  João and Sotiriadis,  Spyros and Guarnieri,  Giacomo and Schmiedmayer,  J\"{o}rg and Eisert,  Jens},
  year = {2025},
  month = jun,
  pages = {1326}
}

@article{Mazzola2009,
  title = {Sudden death and sudden birth of entanglement in common structured reservoirs},
  author = {Mazzola, L. and Maniscalco, S. and Piilo, J. and Suominen, K.-A. and Garraway, B. M.},
  journal = {Phys. Rev. A},
  volume = {79},
  issue = {4},
  pages = {042302},
  numpages = {4},
  year = {2009},
  month = {Apr},
  publisher = {American Physical Society},
  doi = {10.1103/PhysRevA.79.042302},
  url = {https://link.aps.org/doi/10.1103/PhysRevA.79.042302}
}

@article{Sampaio2017,
  title = {Quantifying non-Markovianity due to driving and a finite-size environment in an open quantum system},
  author = {Sampaio, Rui and Suomela, Samu and Schmidt, Rebecca and Ala-Nissila, Tapio},
  journal = {Phys. Rev. A},
  volume = {95},
  issue = {2},
  pages = {022120},
  numpages = {7},
  year = {2017},
  month = {Feb},
  publisher = {American Physical Society},
  doi = {10.1103/PhysRevA.95.022120},
  url = {https://link.aps.org/doi/10.1103/PhysRevA.95.022120}
}

@article{deVega2017,
  title = {Dynamics of non-Markovian open quantum systems},
  author = {de Vega, In\'es and Alonso, Daniel},
  journal = {Rev. Mod. Phys.},
  volume = {89},
  issue = {1},
  pages = {015001},
  numpages = {58},
  year = {2017},
  month = {Jan},
  publisher = {American Physical Society},
  doi = {10.1103/RevModPhys.89.015001},
  url = {https://link.aps.org/doi/10.1103/RevModPhys.89.015001}
}

@article{Lorenzo2011,
  title = {Role of environmental correlations in the non-Markovian dynamics of a spin system},
  author = {Lorenzo, Salvatore and Plastina, Francesco and Paternostro, Mauro},
  journal = {Phys. Rev. A},
  volume = {84},
  issue = {3},
  pages = {032124},
  numpages = {8},
  year = {2011},
  month = {Sep},
  publisher = {American Physical Society},
  doi = {10.1103/PhysRevA.84.032124},
  url = {https://link.aps.org/doi/10.1103/PhysRevA.84.032124}
}

\end{document}